\documentclass[aip, amsmath, amssymb, reprint, 9pt]{revtex4-2}
\usepackage{dcolumn, bm, indentfirst}
\usepackage[title]{appendix}
\usepackage{graphicx, tikz, orcidlink}
\usepackage{epstopdf}
\usepackage{ascmac}
\usepackage{ulem}
\usepackage{cancel}

\usepackage{hyperref}
\hypersetup{colorlinks = true, urlcolor = blue, linkcolor = blue, citecolor = blue}
\begin{document}
\title{The laws of thermodynamics for quantum dissipative systems: A quasi-equilibrium Helmholtz energy approach}

\date{Last updated: \today}

\author{Shoki Koyanagi \orcidlink{0000-0002-8607-1699}}
\author{Yoshitaka Tanimura \orcidlink{0000-0002-7913-054X}}
\email[Author to whom correspondence should be addressed: ]{tanimura.yoshitaka.5w@kyoto-u.jp}
\affiliation{Department of Chemistry, Graduate School of Science, 
Kyoto University, Kyoto 606-8502, Japan}

\begin{abstract}
Using the quasi-equilibrium Helmholtz energy (qHE), defined as the thermodynamic work in a quasi-static process, we investigate the thermal properties of both an isothermal process and a transition process between the adiabatic and isothermal states (adiabatic transition). Here, the work is defined by the change in energy from a steady state to another state under a time-dependent perturbation. In particular, the work for a quasi-static change is regarded as thermodynamic work. We employ a system--bath model that involves time-dependent perturbations in both the system and the system--bath interaction. We conduct numerical experiments for a three-stroke heat machine (a Kelvin-Planck cycle). For this purpose, we employ the hierarchical equations of motion (HEOM) approach. These experiments involve an adiabatic transition field that describes the operation of an adiabatic wall between the system and the bath. Thermodynamic--work diagrams for external fields and their conjugate variables, similar to the $P$--$V$ diagram, are introduced to analyze the work done for the system in the cycle. We find that the thermodynamic efficiency of this machine is zero because the field for the isothermal processes acts as a refrigerator, whereas that for the adiabatic wall acts as a heat engine. This is a numerical manifestation of the Kelvin--Planck statement, which states that it is impossible to derive mechanical effects from a single heat source. These HEOM simulations serve as a rigorous test of thermodynamic formulations because the second law of thermodynamics is only valid when the work involved in the operation of the adiabatic wall is treated accurately.
 \end{abstract}


\maketitle
\section{INTRODUCTION}
\label{sec.intro}

In a process at constant temperature, free energy is an extensive property that describes a mathematically concise relationship between thermodynamic variables under the work done on a system.\cite{Gibbs1876,Gibbs1878,Helmholtz1882} Among free energy functions, Helmholtz energy is an important quantity that links thermodynamics and statistical physics. In the thermodynamics case, Helmholtz energy is defined as the Legendre transform of the internal energy $U$ as $F=U-TS$, where $S$ is entropy, and $T$ is the absolute temperature. In the statistical physics case, Helmholtz energy is evaluated from the partition function $Z$ by $F=-k_B T \ln Z$, where $k_B$ is the Boltzmann constant. Thermodynamic variables such as $U$, $S$, and magnetization $M$ for an external magnetic field $B$ can be obtained from the Helmholtz energy. Note that the free energy defined from the partition function is ubiquitously referred to as the Helmholtz energy. In the thermodynamic sense, the Helmholtz energy for $B$ should be called the Gibbs energy because they appear as the Legendre transformation of $F$ with respect to $B$ and its conjugate property $M$ ($dG = -S dT -M dB$).\cite{Oono2017} Following convention, however, we refer to $F$ as Helmholtz energy.

While investigations based on the partition function are limited in the equilibrium case, the free energy-based investigation has been extensively extended to the non-equilibrium condition, particularly after the discovery of the Jarzynski equality. The non-equilibrium work is related to the equilibrium free energy as $-\ln \left( \langle \exp[-\beta W(t) ] \rangle \right)/\beta =\Delta F_A(t)$.\cite{JarzynskiPRL1997,Jarzynski2004,JarzynskiAnnu11} Here, $\beta \equiv 1/k_{\mathrm{B}}T$ is the inverse temperature, $\Delta F_A(t)$ is the change in the free energy of the system, $W(\tau)$ is the non-equilibrium work, and $\langle \ldots \rangle$ is the ensemble average over all phase-space trajectories under the time-dependent external perturbation from time $t_0$ to $t$. More generally, the fluctuation theorem that can treat non-isothermal cases has been developed.\cite{tasaki2000,kurchan2001quantum,Crooks_2008,YukawaJPSJ00,Mukamel2006FT,Mukamel2019FT,HanggiPRE2007,Hanggi2009FT,Hanggi2011FT,Hanggi2020}

Although investigating such thermodynamic systems in the classical regime is straightforward (e.g., using classical molecular dynamics simulations),\cite{Crooks1999} doing so in the quantum regime remains challenging.\cite{Mukamel2006FT,Mukamel2019FT,HanggiPRE2007,Hanggi2009FT,Hanggi2011FT,Hanggi2020,Segal2006,Thoss2008,YonatanRMP2011,SaitoPRL2013,2015JCao,Alonso2017,WhitneyPhysRevB2018,GalperinJCP2020,SanchezdNatureCarnot2022} For example, the dynamics of a microscopic subsystem (main system) is reversible in time and cannot reach its thermal equilibrium state by itself. Thus, system--bath (SB) models in which a small main system is coupled to a bath are employed to describe the time irreversibility of the dynamics of a reduced system evolving toward the thermal equilibrium state.\cite{Breuer2002,Weiss2012,Binder2018} The temperature of the bath does not change because its heat capacity is infinite; meanwhile, the canonical distribution of the main system on its own cannot be assumed as an equilibrium state due to the presence of the system--bath interaction.\cite{T06JPSJ,T14JCP,T15JCP,T20JCP}

This scenario differs significantly from the classical case where the main system reaches its canonical distribution on its own spontaneously when the system is sufficiently large.
While the Jarzynski equality and fluctuation theorem utilize the ensemble average over all phase-space trajectories under time-dependent external perturbation, such a trajectory-based approach is invalid in a small quantum system because of the measurements of the quantum trajectories.\cite{HanggiPRE2007,ST21JPSJ}

Recently, it was found that the thermodynamic system $A$ coupled to a heat bath $B$ can be described using the quasi-equilibrium Helmholtz energy (qHE) in a physically consistent manner with classical thermodynamics. The qHE is defined as $\Delta F_A(t)=W^{qst} (t)$, where $W^{qst} (t)$ is quasi-static work done on the system during the isothermal operation at time $t$.\cite{ST20JCP,ST21JPSJ} Note that in the present paper, work is defined by the change in energy from one state to another state under a time-dependent perturbation; when the change is quasi-static, it is regarded as thermodynamic work. From $\Delta F_A(t)$, a variety of thermodynamic variables that include the change in internal energy and entropy can be obtained. The previous investigation was limited to the isothermal case;\cite{ST20JCP,ST21JPSJ} in the present study, we extend it to treat thermal transitions between the adiabatic and isothermal states (adiabatic transitions). For this purpose, we introduce a time-dependent SB interaction that describes a manipulation of an adiabatic wall between the system and the bath. 
The key to investigating this quasi-static thermodynamic problem is the HEOM formalism, which enables evaluation of the internal energies of not only the system, but also the bath and the SB interactions, even in low-temperature, non-Markovian, and nonperturbative conditions.\cite{T06JPSJ,T14JCP,T15JCP,KT15JCP,KT16JCP,T20JCP,ST20JCP,ST21JPSJ,Shi2017,ReichmanPhysRevB2013,GoyalKawai2020,Xu_Ankerhold2021}

The qHE treatment for the determination of thermodynamic properties, including entropy change in the adiabatic wall, shall be useful for the study of a quantum heat machine\cite{Nori2007,OttoNori2020, CarnotDeffner2015,Kosloff2014,Kosloff2017,KurizkiReview2015, GalperinPhysRevB2015,GalperinPhysRevB2021,Aurell2017,Carnot_efficiency2018,refrigeSegal2018,PhysRevEOtto2020,2020OttoVictor,Otto2021Venu, OttoNori2020} because the thermodynamic effects of the heat bath play a significant role in an investigation of this kind.\cite{non-MarkovianPezzutto_2019,Non-MarkovGiovannetti2019,Non-MarkovWiedmann_Ankerhold2020,OttoWiedmann_Ankerhold2021,Xu_Ankerhold2022,TensorNetPhysRevX2020,PhysRev2020SBcouple,PhysRevBUncertenty2020} 
In addition, a thermodynamic work diagram of external forces (such as stresses) and their conjugate variables (such as strains), similar to Clapeyron’s $P$--$V$ diagram,\cite{Clapeyron1834} can be introduced to analyze the work done in the system. Such an extension can be useful for analyzing experimental results in the quantum regime, where the quantized work and heat to be manipulated.\cite{JarEXE2015,AdiaEXE2016,KlassRMP2017, LundJosefsson2018,ReimannPRL2018,LundJosefsson2019, nanolettDomenic2019,HuiPRL2020,DEOLIVEIRA2020,Hern_ndez_G_mez_2021}

As a demonstration, we simulate a three-stroke heat machine with a single heat bath that consists of the isothermal process, the adiabatic transition process, and the combination of these processes using the thermodynamic work diagrams. We then show that our numerical results are consistent with the Kelvin--Planck statement (or heat engine statement) of the second law of thermodynamics.\cite{Thomson1851}  The simulations of this cycle can be a critical check on the accuracy of thermodynamic formulations and the description of equations of motion in the quantum regime, since the thermodynamic laws will break down if the work done for the entire system including the adiabatic wall operation is not properly evaluated.

This paper is organized as follows. In Sec. \ref{sec.KIenergy}, we present the definition of the qHE for an arbitrary SB model that satisfies the first and second laws of thermodynamics. In Sec. \ref{sec:model}, we introduce a harmonic oscillator heat-bath and present a spin--Boson model with a time-dependent SB interaction. The HEOM for this system are also introduced.  Section \ref{numericalresults} presents and discusses the numerical results. Finally, concluding remarks are provided in Sec. \ref{conclusion}.

\section{TOTAL ENERGY AND INTERNAL ENERGY}
\label{sec.KIenergy}
\subsection{Total energy}
\label{sec.Kenergy}
We start from a situation where the dynamics of a system are described by a time-dependent Hamiltonian expressed as $\hat H_{tot}(t) $. The total density operator is then expressed as
\begin{align}
\hat \rho_{tot}(t) =&\exp_{+}\left[- \frac{i}{\hbar}\int_{t_0}^t d t' \hat H_{tot}(t') \right] \hat \rho_{tot} (t_0)
\nonumber \\ 
&\times \exp_- \left[ \frac{i}{\hbar} \int_{t_0}^t d t' \hat H_{tot}(t') \right],
\label{eq:rho_S}
\end{align}
where $\exp_{\pm}[ \ldots ]$ are the time-ordered exponentials in which the operators in $[ \ldots ]$ are arranged in chronological order, and $\hat \rho_{tot}(t_0)$ is an initial state normalized as ${\rm{tr}}\{ \hat \rho_{tot}(t_0)\} =1$.
The expectation value for any operator $\hat A$ is defined as $\langle \hat A \rangle = {\rm{tr}}\{\hat A \hat \rho_{tot}(t) \}$. The change in the total energy is expressed as
\begin{align}
\Delta U_{tot} (t) = \langle \hat H_{tot}(t) \rangle - \langle \hat H_{tot}(t_0) \rangle.
\label{each_Energy}
\end{align}
Because we have ${\rm{tr}}\{ [\hat H_{tot} (t), \hat H_{tot}(t)] \hat \rho_{tot}(t)\}=0$, the time derivative of the above is evaluated as
\begin{align}
\frac{d}{dt} \Delta U_{tot} (t) =P_{tot} (t),
\label{TDO2}
\end{align}
where $P_{tot} (t) \equiv {\rm{tr}} \{ (\partial {\hat H_{tot} (t)}/{ \partial t}) \hat \rho_{tot}(t) \}$. 
Thus, the conservation law of the total energy is expressed as
\begin{align}
\Delta U_{tot} (t) = W_{tot} (t),
\label{DU_tot_Wtot}
\end{align}
where 
\begin{align}
W_{tot} (t) = \int_{t_0}^t dt' {\rm{tr}}\left\{ \frac{\partial \hat H_{tot} (t')}{ \partial t'} \hat \rho_{tot}(t') \right\}.
\label{work}
\end{align}
The work here is defined as the change in energy from one state to another under a time-dependent perturbation. This work is attributed to external perturbations. The above equality represents the relationship between the work and the total energy in any Hamiltonian $\hat H_{tot} (t)$ and is not restricted to thermodynamic systems characterized by temperature.

\subsection{Internal energy}
\label{sec.Ienergy}
To introduce thermodynamic temperature in the framework of open quantum dynamics theory, we next consider a quasi-static process in which the system changes slowly under weak external perturbation while maintaining a quasi-equilibrium state. 
Thus, we assume that the total system is described by the quasi-equilibrium partition function (qPF), which is defined as $Z_{tot} (t) = {\operatorname{tr}} \{ \exp[-\beta {\hat H_{tot}}(t) ] \} $.\cite{YukawaJPSJ00,Mukamel2006FT,Mukamel2019FT,HanggiPRE2007,Hanggi2009FT,Hanggi2011FT,Hanggi2020} We then have
\begin{align}
\frac{\partial \ln Z_{tot} (t) }{\partial t} 
= - \beta {\operatorname{tr}} \left\{ \frac{\partial {\hat H_{tot}}(t)} {\partial t} {\hat \rho}_{tot} (\beta; t) \right\},
\label{eq:ZADerivative}
\end{align} 
where ${\hat \rho}_{tot} (\beta; t) \equiv \exp[-\beta {\hat H_{tot}}(t)]/Z_{tot}(t)$. Thus, we define
\begin{align}
\ln \left( \frac{Z_{tot} (t)}{Z_{tot} (t_0)} \right) =- \beta \Delta F_{tot} (\beta; t),
\label{Z_tot01} 
\end{align}
where 
\begin{align}
\Delta F_{tot} (\beta; t) \equiv \int_{t_0}^t dt' {\operatorname{tr}} \left\{ \frac{\partial {\hat H_{tot}(t')}}{\partial t'} {\hat \rho}_{tot} (\beta; t') \right\} 
\label{F_tot2} 
\end{align}
is the qHE.\cite{ST20JCP,ST21JPSJ} The significant difference between the qHE and conventional free energy is that the qHE involves time-dependent work done for the system and is thus a time-dependent function. Therefore, the qHE vanishes when there is no work. 

For the inverse bath temperature $\beta$, the qHE and the work defined by Eq. \eqref{work} satisfy the minimum work principle expressed as\cite{Oono2017,Lenard1978,tasaki2000}
\begin{align}
W_{tot} (\beta; t) \ge \Delta F_{tot} (\beta; t),
\label{Ineq} 
\end{align}
which corresponds to the second law of thermodynamics. Although the above inequality has been derived for an isolated quantum system, it can also be applied to a SB system, because the total system of a SB model is regarded as an isolated system. (See Appendix~\ref{AppQHE}). As we will show below, we can utilize the above inequality for open quantum dynamics systems, because the HEOM formalism allows us to evaluate the work and free energy not only for a reduced system but also a bath. 

In general, work is not a state variable, but as has been shown, it is the state variable in quasi-static processes where the equality sign holds.\cite{ST20JCP,ST21JPSJ}
Using the qPF-based density operator, we can evaluate the change in the total internal energy as
\begin{align}
\Delta U_{tot} (\beta; t) &= \mathrm{tr} \left\{ \hat H_{tot}(t) {\hat \rho}_{tot} (\beta; t) \right\} \nonumber \\
&-\mathrm{tr} \left\{ \hat H_{tot} (t_0) {\hat \rho}_{tot} (\beta; t_0) \right\},
\label{each_internalEnergy}
\end{align}
which agrees with $\Delta U_{tot} (\beta; t)= \partial ( \beta \Delta F_{tot} (\beta; t) ) / \partial \beta$. We can also evaluate the change in heat as $Q_{tot} (\beta; t)=\beta \partial \Delta F_{tot} (\beta; t) / \partial \beta$. 
From the above relations, the first law of thermodynamics can be expressed as
\begin{align}
\Delta U_{tot} (\beta; t) = Q_{tot} (\beta; t) + W_{tot} (\beta ; t),
\label{firstlaw}
\end{align}
where we used $W_{tot} ( \beta ; t) \equiv \Delta F_{tot}(\beta; t)$ from the definition in Eq. \eqref{work}.
From Eq. \eqref{DU_tot_Wtot} under the quasi-equilibrium condition, we have 
\begin{align}
\Delta U_{tot}^{qeq} (\beta ; t) = \Delta U_{tot} (\beta; t) - Q_{tot} (\beta; t),
\end{align}
which indicates that the difference between the system energy and the internal energy is the heat $Q_{tot} (\beta; t)$.

\section{OPEN QUANTUM DYNAMICS THEORY FOR THERMODYNAMICS}
\label{sec:model}
\subsection{System--bath model for isothermal--adiabatic transitions}

To proceed one step, we consider a SB model to compute the thermodynamic variables. 
The total Hamiltonian is expressed as $\hat H_{tot}(t) = \hat H_A(t) + \hat H_I(t) + \hat H_B$, where $\hat H_A (t)$ and $\hat H_I(t)$ are the time-dependent Hamiltonians of the main system and the SB interaction, respectively, and $\hat H_B$ is the Hamiltonian of the bath. The time dependences of $\hat H_A(t)$ and $\hat H_I(t)$ are respectively described by the isothermal driving field (IDF) and adiabatic transition field (ATF), which are represented by $B(t)$ and $A(t)$, respectively. By choosing $B(t)$ and $A(t)$, a real experimental situation can be simulated.\cite{JarEXE2015,AdiaEXE2016,KlassRMP2017, LundJosefsson2018,ReimannPRL2018, LundJosefsson2019, nanolettDomenic2019,HuiPRL2020,DEOLIVEIRA2020,Hern_ndez_G_mez_2021} Later, IDF and ATF are given explicit forms when conducting numerical simulations. 

We employ a heat bath modeled by an ensemble of harmonic oscillators:
\begin{align}
\hat H_B = \sum\limits_{j=1}^{N} {\left( \frac{\hat p_j^2 }{2m_j } + \frac{1}{2}m_j \omega _j^2 \hat x_j^2 \right), }
\label{Bath}
\end{align}
with the momentum, position, mass, and frequency of the $j$th bath oscillator given by $\hat{p}_{j}$, $\hat{x}_{j}$, $m_{j}$, and $\omega_{j}$, respectively. To describe the transition between the isothermal and adiabatic processes, we consider the SB interaction expressed as
\begin{align}
\hat H_I(t) = A (t) \hat V\sum\limits_{j=1}^{N} {c_j \hat x_j},
\label{SBI}
\end{align}
where $\hat V$ is the system part of the interaction, and $c_j$ is the $j$th coupling constant. With a proper choice of $A(t)$, $\hat V$, and the spectral distribution of the bath coupling, a variety of isothermal--adiabatic manipulations (e.g., the insertion and exertion of the adiabatic wall or attaching or removing the quantum system to the bath) can be performed. 

The open quantum dynamics theory utilizes the reduced density operator (RDO) assuming that the heat bath is in the thermal equilibrium state at $\beta$. When the bath part of the SB interaction is a linear function of the bath coordinates, as in Eq. \eqref{SBI}, we can eliminate the bath degrees of freedom by performing the Gaussian integrations involved in the bath Hamiltonian. This leads to the following reduced description of the system operator:
\begin{eqnarray}
\hat{\rho}_{A}^{rd}(t) = \frac{{\rm{tr}}_B \{ \hat{\rho}_{tot} (t) \}}{Z_B^0(\beta)},
\label{reduced}
\end{eqnarray}
where the denominator $Z_{B}^{0}(\beta) = {\rm tr}_{B} \{ \exp[-\beta  \hat H_{B} ] \}$ is introduced to maintain the reduced operator in a finite value, and the RDO is normalized as $\rm{tr}_A \{\hat{\rho}_{A}^{rd}(t) \}=1$.

Due to the Bosonic nature of the bath, all bath effects on the system are determined by the bath correlation function
$C(t) \equiv \langle \hat{X}(t) \hat{X}(0) \rangle_\mathrm{B}$, where $\hat{X} \equiv \sum_j c_{j} \hat{x}_{j} $ is the collective coordinate of the bath, and $\langle \ldots \rangle_\mathrm{B}$ represents the average taken with respect to the canonical density operator of the bath.
The bath correlation function is expressed in terms of the bath spectral density $J(\omega)$ as
\begin{align}
C(t)
= \int_0^\infty d\omega \, J(\omega)
\left[ \coth\left( \frac{\beta \hbar\omega}{2} \right) \cos(\omega t)
- i \sin(\omega t) \right],
\label{eq:BCF}
\end{align}
where $J(\omega) \equiv \sum_{j=1}^{N} ({\hbar c_{j}^2}/2 m_{j} \omega_{j}) \delta ( \omega - \omega_{j} )$.
The real part of Eq.~(\ref{eq:BCF}) is analogous to the classical correlation function of the bath and represents the fluctuations, while its imaginary part represents the dissipation. The fluctuation term is related to the dissipation term through the quantum version of the fluctuation---dissipation theorem.\cite{TK89JPSJ1, T06JPSJ,T20JCP}
For the heat bath to be an unlimited heat source with an infinite heat capacity, the number of heat bath oscillators $N$ can be made infinitely large by replacing $J (\omega)$ with a continuous distribution.

We should note that in the framework of regular open quantum dynamic theories, the energy conservation law in Eq.~\eqref{DU_tot_Wtot} does not hold due to the reduced description of the density operator $\hat{\rho}_{tot} (t)$. For example, without the external force (i.e., $\hat H_{tot}(t)=\hat H_{tot}$), the total system reaches the equilibrium state $\hat \rho_{tot}^{eq}=\exp[-\beta \hat H_{tot}]$ for sufficiently long $t$, even if we start from a non-equilibrium initial condition expressed as $\hat \rho_{tot}(t_0) ={\hat \rho}_A(t_0) \exp[-\beta \hat H_B ]$, where ${\hat \rho}_A(t_0)$ is a highly excited initial state of the system. This indicates that we have $\Delta U_{tot} (t) \ne 0$ for a dissipative system; in contrast, from Eq.~\eqref{each_Energy}, we have $\Delta U_{tot} (t) = 0$ for the total system.
In open quantum dynamics theory, this phenomenon arises because we have reduced the heat-bath degrees of freedom prior to the total system evolving in time. Thus, to apply the qPF theory introduced in Sec. \ref{sec.Ienergy} to the SB model, we must extend the reduced dynamics theory to separately evaluate the energy flow to the bath.

\subsection{HEOM}

We now introduce the HEOM formalism that plays a key role in the present quantum thermodynamic investigations.
In the HEOM formalism, the set of equations of motion consists of the auxiliary density operators (ADOs).\cite{TK89JPSJ1,T90PRA,IT05JPSJ,T06JPSJ,T14JCP,T15JCP,KT15JCP,KT16JCP,T20JCP,ST20JCP,ST21JPSJ}
Here, we consider the case that the bath correlation function [Eq.~\eqref{eq:BCF}] is written as a linear combination of exponential functions, $C(t) = \sum_{l=0}^{K} \zeta_{l} e^{-\nu_{l} |t|}$, where $\nu_l$ and $\zeta_l$ are the respectively frequency and strength obtained from a Pad{\'e} spectral decomposition scheme to reduce the hierarchy size.\cite{hu2010communication} The ADOs introduced in the HEOM are defined 
by the index 
$\vec{n} = ( n_{0} , n_{1} , \cdots , n_{K} )$, where $n_{l}$ takes an integer value zero and above. The zeroth ADO for which all elements are zero, $\vec{0} = ( 0 , 0 , \cdots , 0 )$,
corresponds to the actual RDO, $\hat \rho_A^{rd} (t) =\hat{\rho}_{\vec{0}} ( t )$, and we normalize the RDO as $\mathrm{tr}_A \{ \hat{\rho}_{\vec{0}} ( t ) \} =1$.
 The SB coupling strength of the equations of motion for the ADOs depends on time. The HEOM is then expressed as
\begin{align}
& \frac{\partial}{\partial t} \hat{\rho}_{\vec{n}} ( t ) = \left( - \frac{i}{\hbar} \hat{H}_{\mathrm{A}}^\times ( t ) - \sum_{l = 0}^{K}n_{l} \nu_{l} \right) \hat{\rho}_{\vec{n}} ( t ) \nonumber \\
& \quad - \frac{i A( t )}{\hbar} \sum_{l = 0}^{K} n_{l} \hat{\Theta}_{l} \hat{\rho}_{\vec{n} - \vec{e}_l} ( t ) - \frac{i A( t )}{\hbar} \hat{V}^\times \sum_{l = 0}^K 
\hat{\rho}_{\vec{n} + \vec{e}_l} ( t ),
\label{ModelHEOM}
\end{align}
where $\vec{e}_{l}$ is the ($K$+1)-dimensional unit vector. The operators are defined as
\begin{equation}
\hat{\Theta}_{0} = \left( \frac{\gamma}{\beta} + \sum_{l = 1}^K 
\frac{\zeta_l}{ \gamma^2}{\beta} \frac{2 \gamma}{\gamma^2 - \nu_{l}^2} \right)
\hat{V}^\times - \frac{i \hbar \gamma^2}{2} \hat{V}^\circ,
\end{equation}
and
\begin{equation}
\hat{\Theta}_{l} = - \frac{ \zeta_{l} \gamma^2}{\beta}
\frac{2 \nu_{l}}{\gamma^2 - \nu_{l}^2} \hat{V}^\times
\qquad ( l \in \{ 1 , 2, \cdots , K \} ),
\end{equation}
where $\hat{\mathcal{O}}^\times \hat{\mathcal{P}} = [ \hat{\mathcal{O}} , \hat{\mathcal{P}} ]$ and
$\hat{\mathcal{O}}^\circ \hat{\mathcal{P}} = \{ \hat{\mathcal{O}} , \hat{\mathcal{P}} \}$ for arbitrary
operators $\hat{\mathcal{O}}$ and $\hat{\mathcal{P}}$.

In principle, the HEOM provides an asymptotic approach to calculate various physical quantities with any desired accuracy by adjusting the number of hierarchal elements determined by $K$; the error introduced by the truncation is negligibly small in the case that $K$ is sufficiently large.\cite{T20JCP}

\subsection{Physical variables}
\subsubsection{System energy and work}

We first evaluate the change in the energy of each part of the Hamiltonian, defined as $U_{\alpha} (t)= {\rm{tr}} \{{\hat H_{\alpha}}(t) \hat \rho_{tot}(t)\}$ for $\alpha= A, I,$ and $B$, where $ \hat \rho_{tot}(t)$ is given by Eq.~\eqref{eq:rho_S}. 

Although the evaluations of $U_{I} (t)$ and $U_{B} (t)$ are not easy within the framework of the open quantum dynamics theory because the bath degrees of freedom have been reduced, we can obtain their values indirectly using the hierarchical elements in the HEOM formalism. This is because in the HEOM formalism, the higher hierarchical elements store information about the higher cumulant of the bath coordinates, as previously demonstrated.\cite{KT15JCP,KT16JCP,ST20JCP,ST21JPSJ,Shi2017}

We express the HEOM elements obtained from Eq.~\eqref{ModelHEOM} under any form of external field as $ \hat{\rho}_{\vec{n}}( t )$. Using the zeroth member of the hierarchy $\hat{\rho}_{\vec{0}} ( t ) $, the expectation value of the system energy at time $t$ is evaluated as
\begin{align}
U_{A} (t) = \mathrm{tr}_A \left\{ \hat{H}_{\mathrm{A}} ( t ) \hat{\rho}_{\vec{0}} ( t ) \right\}.
\label{expectHA}
\end{align} 
In the HEOM formalism, the first-order hierarchical elements $\hat{\rho}_{\vec{e}_l} ( t )$ ($0 \le l \le K$) are defined as the expectation value of the collective bath coordinate $\hat X$. Thus, from Eq. \eqref {SBI}, the SB interaction energy is expressed as\cite{KT15JCP,KT16JCP,ST20JCP,ST21JPSJ}
\begin{align}
U_{I} (t)= A ( t ) \sum_{l = 0}^{K} \mathrm{tr}_A \left\{ \hat{V} \hat{\rho}_{\vec{e}_l} ( t ) \right\},
\label{expectHI}
\end{align}
where $\vec{e}_l$ is the index for the first-order hierarchical member. To evaluate the bath energy, we consider the expectation value of bath energy defined as $U_{B} (t)\equiv {\rm{tr}}\{   { \hat H}_{B} {\hat \rho}_{tot}(t)\}$ and evaluate the change in the bath energy from $\partial U_{B} (t) / \partial t = i \int_{t_0}^{t} {\rm{tr}}\{ [\hat H_{I}, \hat H_{B}] \hat \rho_{tot}(t) \} dt/\hbar$, which is obtained from $\partial \hat{\rho}_{tot} ( t ) / \partial t = [ \hat{H}_{tot} ( t ) , \hat{\rho}_{tot} ( t ) ] / i \hbar$. The bath energy is then evaluated as (see Appendix \ref{AppExpectHB})
\begin{align}
\frac{\partial}{\partial t}
U_{B} (t)&= A ( t ) \sum_{l = 0}^{K} \nu_{l}
\mathrm{tr}_{A} \{ \hat{V} \hat{\rho}_{\vec{e}_l} ( t ) \} \nonumber \\
& + A^2 ( t ) \gamma^2 \mathrm{tr}_A \left\{ \hat V^2 \hat{\rho}_{\vec{0}} ( t ) \right\}.
\label{expectHB}
\end{align}
The total work is evaluated as
\begin{align}
W_{tot} (t) = W_{A} (t) + W_{I} (t),
\label{workWtot}
\end{align}
where
\begin{align}
W_{A} (t) = \int_{t_0}^t dt' \mathrm{tr}_A \left\{ \frac{\partial {\hat H}_A( t' )}{\partial t} \hat{\rho}_{\vec{0}} (t') \right\},
\label{workWA}
\end{align}
and
\begin{align}
W_{I} (t) = \int_{t_0}^t dt'\frac{d A (t')}{d {t'}} \sum_{l = 0}^K \mathrm{tr}_A \{ \hat{V} \hat{\rho}_{\vec{e}_l } ( t' ) \}.
\label{workWI}
\end{align}
We thus have $\Delta U_{tot} (t) = \Delta U_{A} (t) + \Delta U_{I} (t)+ \Delta U_{B} (t)$, where $\Delta U_{\alpha} (t) \equiv U_{\alpha} (t) - U_{\alpha} (t_0)$ for $\alpha=A, B, I$, and $tot$. The change in total energy and the work evaluated from the HEOM now satisfy the energy conservation law in Eq.~\eqref{DU_tot_Wtot}.

\subsubsection{Free energies and partition functions}

For the SB Hamiltonian, the RDO is expressed as Eq.~\eqref{reduced}.  By using the above definition with the HEOM, the work and internal energy changes are calculated numerically and rigorously for any thermal cycle driven by the IDF and ATF regardless of the condition in Eq. \eqref{Ineq}. Nevertheless, we limit our discussion to the quasi-static case and attempt to quantify quantum thermodynamic variables by comparing physical quantities calculated in the HEOM with thermodynamic quantities evaluated in the qHE formalism. Then, the reduced qPF operator is assumed to be of the form ${\hat Z}_{A + I}^{rd} (\beta; t) = \operatorname{tr}_B \{\exp[-\beta \hat H_{tot}(t) ] \}/Z_B^0 (\beta)$, which has been evaluated as ${\hat Z}_{A + I}^{rd} (\beta; t)= \exp[ -\beta {\hat H}^\ast (t) ]$, where  ${\hat H}^\ast (t)$ is referred to as the Hamiltonian of mean force or effective Hamiltonian.\cite{Hanggi2009FT,Hanggi2011FT,Hanggi2020} We then have $Z_{tot} (\beta; t) = Z_{A + I}^{rd}(\beta; t) Z_B^0 (\beta)$, where $ Z_{A + I}^{rd}(\beta ; t)=\operatorname{tr}_A \{{\hat Z}_{A+ I}^{rd} (\beta; t)\}$. 
In this study, we evaluate $Z_{tot}$ from the HEOM approach, including a contribution from SB coupling using Eq.~\eqref{workWI}. We define the RDO for the PF as ${\hat \rho}_{tot} (\beta; t)\equiv \exp[-\beta \hat H_{tot}(t) ]/Z_{tot} (\beta; t)$ and ${\hat \rho}_{A}^{rd} (\beta; t) \equiv \operatorname{tr}_B \{\exp[-\beta \hat H_{tot}(t) ] \}/Z_{tot} (\beta; t)$. 
Accordingly, we evaluate Eq.~\eqref{Z_tot01} as 
\begin{align}
\ln \left( \frac{Z_{A+I}^{rd} (\beta ; t)}{Z_{A+I}^{rd} (\beta ; t_0)} \right) =- \beta \Delta F_{A + I}^{rd}(\beta; t ),
\label{Z_totR01} 
\end{align}
where $\Delta F_{A + I}^{rd}(\beta; t )$ is the reduced free energy. From Eqs.~\eqref{workWA} and \eqref{workWI}, we have
\begin{equation}
\Delta F_{A+I}^{rd} (\beta; t ) = \Delta F_A^{qst} (\beta; t ) + \Delta F_I^{qst} (\beta; t ),
\label{eq:Ftotqeq}
\end{equation}
where
\begin{align}
\Delta F_A^{qst} (\beta; t ) = \int_{t_0}^t dt' \mathrm{tr}_A \left\{ \frac{\partial {\hat H}_A( t' )}{\partial t} \hat{\rho}_{\vec{0}}^{qst} (t') \right\},
\label{qHEWA}
\end{align}
and
\begin{align}
\Delta F_I^{qst} (\beta; t ) = \int_{t_0}^t dt'\frac{d A (t')}{d {t'}} \sum_{l = 0}^K \mathrm{tr}_A \{ \hat{V} \hat{\rho}_{\vec{e}_l }^{qst} ( t' ) \}.
\label{qHEWI}
\end{align}
From the above, we define $\Delta U_{\alpha}^{rd} (\beta; t) \equiv \partial ( \beta \Delta F_{\alpha}^{qst}(\beta; t) ) / \partial \beta$ and $\Delta S_{\alpha}^{rd} (\beta; t)\equiv k_B \beta^2 \partial \Delta F_{\alpha}^{qst} (\beta; t) / \partial \beta$ for $\alpha=A$ and $I$. Note that we introduced the suffix {\it rd} in addition to {\it qst}, because $\Delta U_{A}^{rd} (\beta; t)$ involves the contribution form the system part of the SB interaction, as we briefly explain in Sec. \ref{sec.ReducedH}.\cite{ST20JCP}
Using the qHE, we can introduce the conjugate variables of the IDF and ATF as
\begin{equation}
M (\beta; t) \equiv - \frac{\partial \Delta F_{A}^{qst}(\beta; t)}{\partial {B(t)}}, 
\end{equation}
and
\begin{equation}
D (\beta; t) \equiv - \frac{\partial \Delta F_{I}^{qst}(\beta; t)}{\partial {A(t)}},
\end{equation}
where $B(t)$ and $A(t)$ represent, for example, the magnetic field and stress, whereas $M(\beta; t)$ and $D(\beta; t)$ represent, for example, the magnetization and strain.
From the definition of the total work Eq.\eqref{work}, we can evaluate the above variables in terms of the  ADOs as
\begin{equation}
M (\beta; t)  = \mathrm{tr}_A \left\{ \hat{\sigma}_z \hat{\rho}_{\vec{0}} ( t ) \right\},
\end{equation}
and
\begin{equation}
D (\beta; t) 
= - \sum_{l = 0}^{K_k} \mathrm{tr}_A \left\{ \hat{V} \hat{\rho}_{\vec{e}^k_l} ( t ) \right\}.
\end{equation}
Note that $M(\beta; t)$ and $D(\beta; t)$ are state variables in the quasi-static case, because they are uniquely determined by the state specified by the quasi-equilibrium distribution at $t$ and are independent of the pathway of work. As described in Eq.~\eqref{firstlaw}, we have the first law of thermodynamics for each component $\alpha=A$ and $I$ as
\begin{align}
\Delta U^{rd}_{\alpha} (\beta; t) = T \Delta S_{\alpha}^{rd} (\beta; t) + W_{\alpha}^{qst}(\beta ; t),
\label{firstlaw1}
\end{align}
where 
\begin{align}
W_{A}^{qst} (\beta ; t) = - \int_{t_0}^t \frac{d B ( t' )}{d t'} M(\beta; t') d t',
\label{WAqst}
\end{align}
and
\begin{align}
W_{I}^{qst} (\beta ; t) = - \int_{t_0}^t \frac{d A ( t' )}{d t'} D(\beta; t') d t'.
\label{WIqst}
\end{align}

For $\Delta U_{A+I}^{rd} (\beta; t)=\Delta U^{rd}_{A} (\beta; t) +\Delta U^{rd}_{I} (\beta; t)$, $Q_{A+I}^{rd}(\beta; t)= T  \Delta S^{rd}_{A+I}(\beta; t)$ with $\Delta S_{A+I}^{rd}(\beta; t)= \Delta S^{rd}_{A} (\beta; t)+ \Delta S^{rd}_{I} (\beta; t)$, and $ W_{tot} (\beta ; t)= W_{A}^{qst}(\beta ; t)+ W_{I}^{qst}(\beta ; t)$, we have 
\begin{align}
\Delta U_{A+I}^{rd} (\beta; t)= Q_{A+I}^{rd}(\beta; t) + W_{tot} (\beta ; t).
\label{firstlaw3}
\end{align}
Here, work is defined by the quasi-static change in total energy under time-dependent perturbation. This work is regarded as thermodynamic work. 
When the main system consists of $n$ non-interacting spins that are independently coupled to the heat bath, the magnitude of $\Delta U_{A+I}^{rd}(\beta; t)$, $Q_{A+I}^{rd}(\beta; t)$, $\Delta S_{A+I}^{rd}(\beta ; t)$, $ W_{tot} (\beta ; t)$, $M(\beta; t) $, and $D(\beta; t) $ are proportional to $n$. Thus, they are extensive properties, whereas $B(t)$, $A(t)$, and $T$ are intensive properties. 

Using the qHE, we can also define the reduced PF for $\alpha = A$ and $I$ as
\begin{align}
Z_{\alpha}^{qst} (\beta ; t) = \exp[ - \beta \Delta F_{\alpha}^{qst}(\beta; t)] Z_{\alpha}^{qst} (\beta ; t_0).
\label{Zpart} 
\end{align}
The total PF is then expressed as $Z_{tot}(\beta ; t) = Z_{A+I}^{rd} (\beta ; t) Z_B^0 (\beta)$, where $Z_{A+I}^{rd}( \beta ; t) = Z_{A}^{qst} ( \beta ; t) Z_{I}^{qst} ( \beta ; t)$, which is consistent with Eq.~\eqref{firstlaw3} for $W_{tot}(t)=\Delta F_{A+I}^{rd}(\beta; t)$.
However, from the above, we have $\Delta U_{tot} (\beta; t)=\Delta U_{A+I}^{rd} (\beta; t)$ and $Q_{tot} (\beta; t)=Q_{A+I}^{rd}(\beta; t)$, which contradicts the first law of thermodynamics for the total system presented in Eq.~\eqref{firstlaw}. 
To illustrate this point, we consider the change in the internal energy of each Hamiltonian component defined as
\begin{align}
\Delta U_{\alpha}^{qst} (\beta; t) \equiv U_{\alpha}^{qst} (\beta; t) - U_{\alpha}^{qst} (\beta; t_0),
\label{InternalEI}
\end{align}
where $ U_{\alpha}^{qst} (\beta; t)$ is evaluated from Eqs.~\eqref{expectHA}--\eqref{expectHB} with the use of $\hat{\rho}_{\vec{n}}^{qst} (t)$ for $\alpha=A, B$, and $I$. The total internal energy is then given by
\begin{align}
\Delta U_{tot} (\beta; t) =\Delta U_{A}^{qst} (\beta; t) +\Delta U_{I}^{qst} (\beta; t) +\Delta U_{B}^{qst} (\beta; t),
\label{Internaltot}
\end{align}
whereas we obtain $\Delta U_{tot}^{rd} (\beta; t)=\Delta U_{A}^{rd} (\beta; t)+\Delta U_{I}^{rd} (\beta; t)$ from $Z_{tot}(\beta ; t) = Z_{A+I}^{rd} (\beta ; t) Z_B^0(\beta)$, where $\Delta U^{rd}_{tot} ( \beta ; t ) = - \partial \ln Z_{tot} ( \beta ; t ) / \partial \beta$ . What is missing here is the change in the bath internal energy evaluated from Eq.~\eqref{expectHB}. This difference arises because the reduced description of the system cannot evaluate the change in the bath energy, whereas the HEOM formalism can include the change in bath energy from Eq.~\eqref{expectHB}. 

\subsection{Reduced heat-bath energy}
\label{sec.ReducedH}
To illustrate the above point, we consider the isothermal and the adiabatic transition processes for $B(t)$ and $A(t)$.\cite{ST20JCP,ST21JPSJ} We then have $Z_{tot}(\beta; t)=Z_{A}^{qst} (\beta; t) Z_{I}^{qst} ( \beta ; t ) Z_{B}^{0}(\beta)$, which leads to $\Delta U_{tot}(\beta; t) = \Delta U_{A + I}^{rd}(\beta; t)$ when we differentiate both sides with regard to $\beta$.  This indicates that there is no heat flow between the system and the bath, due to the constraints of the reduced description of the system, even when we consider the non-equilibrium situation. Then, as illustrated in Eq.~\eqref{Internaltot}, we must compensate for the change in the SB interaction energy using the HEOM formalism.

It is important to note that we have $| \Delta U_{A + I}^{rd}(\beta; t) - \Delta U_{A + I}^{qst}(\beta; t) | \neq 0$ because $\Delta U_{A + I}^{rd}(\beta; t)$ doesn't include the contribution from the bath part of the SB interaction due to the reduced description of the system.\cite{ST20JCP} We then separate the system part of the SB interaction as $\Delta U_I^{A}(\beta; t) \equiv \Delta U_{A + I}^{rd}(\beta; t) - \Delta U_{A}^{qst}(\beta; t)$. The bath part of the internal energy in the SB interaction is expressed as $\Delta U_I^{B}(\beta; t) \equiv \Delta U_I^{qst}(\beta; t) -\Delta U_I^{A}(\beta; t)$. Thus, we have the change in the internal energy of the reduced bath system as
\begin{align} 
\Delta U_{B}^{rd}(\beta; t) = \Delta U_I^{B}(\beta; t) + \Delta U_{B}^{qst}(\beta; t).
\label{U_Brd} 
\end{align}
Because there is no external force on the bath, the qHE for the reduced bath vanishes [i.e., $\ln \left( Z_{B}^{rd} (t)/Z_{B}^{rd} (t_0) \right) = 0$]. Thus, we have
\begin{align} 
\Delta U_{B}^{rd}(\beta; t) = Q_{B}^{rd}(\beta; t).
\label{U_Brd2} 
\end{align}
Since $Q_{tot} ( \beta ; t ) = Q^{rd}_{A + I} ( \beta ; t ) + Q^{rd}_B ( \beta ; t ) = 0$, the first law of thermodynamics for the total system is expressed as 
\begin{align}
\Delta U_{tot} (\beta; t)=  W_{tot} (t),
\label{firstlaw4}
\end{align}
where $\Delta U_{tot} (\beta; t)= \Delta U_{A + I}^{rd}(\beta; t)+ \Delta U_{B}^{rd}(\beta; t)$, and $W_{tot} (\beta ; t)= W_{A + I}^{qst}(\beta ; t)= \Delta F_{A + I}^{rd} (\beta; t )$.
Although a similar expression has been employed in a perturbative Markovian case in which $\Delta U_{B}^{rd}(\beta; t ) \approx \Delta U_{B}(\beta; t)$,\cite{Mukamel2006FT,Mukamel2019FT}  the present expression is valid even in a non-Markovian and nonperturbative case. For the bath part of heat $Q_B(\beta ; t) = \Delta U_I^{B}(\beta; t) + \Delta U_{B}(\beta; t) $, we thus obtain the total entropy production as
\begin{align} 
\Sigma _{tot} (\beta ; t) = \Delta S_A^{rd} (\beta ; t) + \beta Q_{B}^{rd}(\beta; t).
\label{Sigma_tot} 
\end{align}
It should be noted that when considering the thermodynamic properties of $A$, we can ignore the effects of the bath.

Note that for the situation that describes a thermal transition from an adiabatic state to an isothermal state described by the fixed IDF $B(t)=B_0$ with the ATF $A(t)$, 
the work $W_{I}^{qst} (\beta ; t)$ of inserting or removing the adiabatic wall that applies to both the system and the bath plays a role.
The heat $Q_{tot}(\beta; t)$ is generated during the manipulation of the adiabatic wall, indicating that Maxwell's demon\cite{ Maxwelldemon2006,Maxwelldemon2009} for thermal processes also obeys the thermodynamic law.

\section{NUMERICAL RESULTS}
\label{numericalresults}
\renewcommand{\arraystretch}{1.3}
\begin{table}[h]
\caption{\label{Cycelarameter}Time evolutions of IDF [$B(t)$] and ATF [$A(t)$] for a three-stroke heat machine (a Kelvin-Planck cycle) with equal time intervals $\tau$. The cycle consists of (i) isothermal expansion, (ii) adiabatic transition, and (iii) the combination of isothermal compression and diabatic transition.}
\centering
\begin{tabular}{|c|c|c|}
\hline
&   $B( t )/|B_0|$ & $A( t )/|A_0| $
\\
\hline
%& $t < 0$ & $1.0$ & $1.0$
%\\
%\hline
(i)  & $1.0 + t / \tau $ & $1.0$
\\
\hline
(ii)  & $2.0$ & $1.3 - 0.3t / \tau $
\\
\hline
(iii)  & $4.0 - t  /\tau $ & $0.1 + 0.3 t / \tau$
\\
%\hline
%& $T < t$ & $1.0$ & $1.0$
%\\
\hline
\end{tabular}
\end{table}
\renewcommand{\arraystretch}{1.0}
\begin{figure}
\centering
\includegraphics[width=8cm]{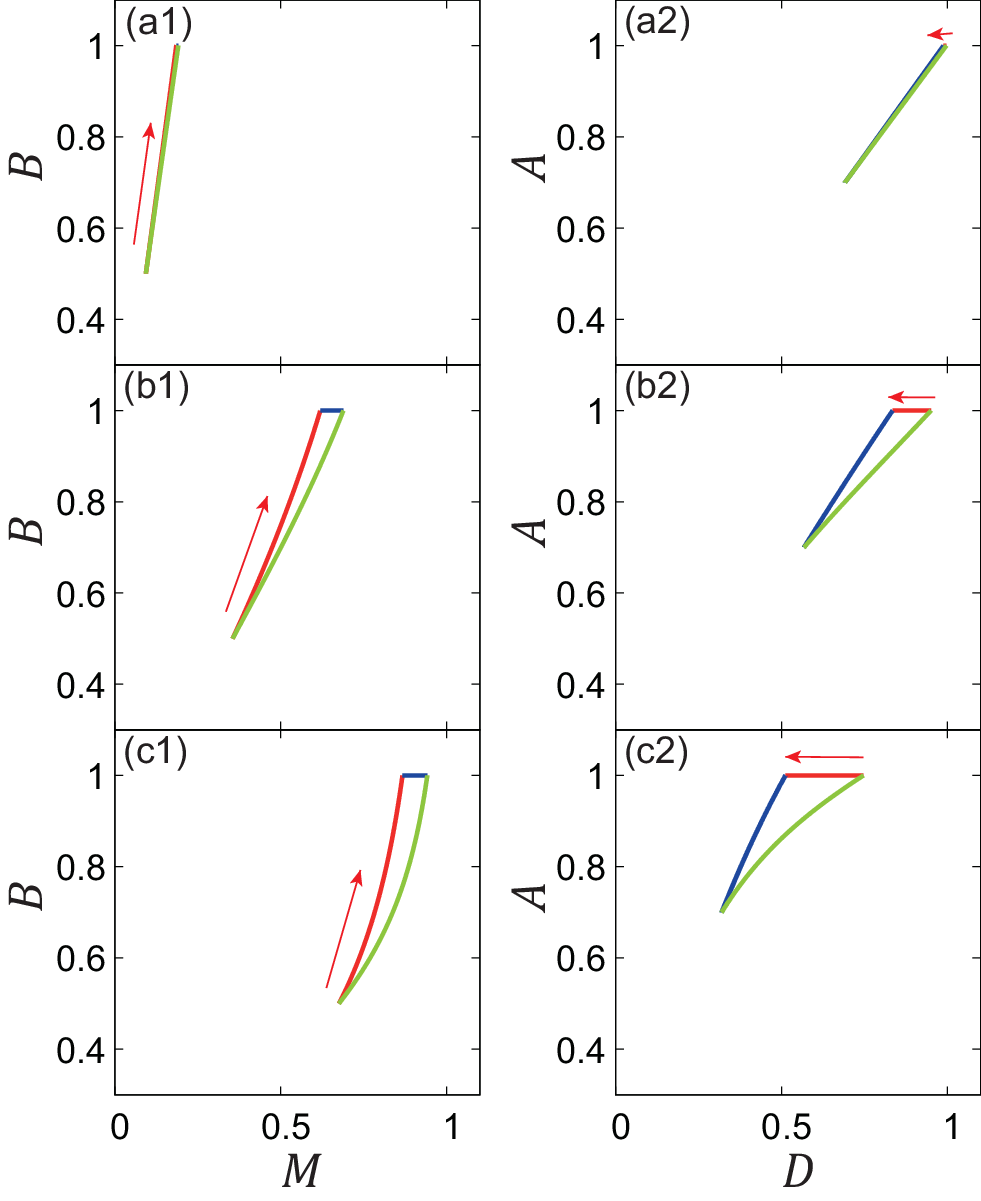}
\caption{\label{BMADDiagram} $B$--$M$ diagrams (left column) and $A$--$D$ diagrams (right column) for the three-stroke heat-machine (a Kelvin-Planck cycle) described by the ATF [$A(t)$] and IDF [$B(t)$] in Table~\ref{Cycelarameter} at different temperatures: (a) $\beta \hbar =0.2$ (high), (b) $\beta \hbar = 1.0$ (intermediate), and (c) $\beta \hbar = 5.0$ (low). In each figure, the cycle starts from the red arrow, and the three curves represent (i) isothermal expansion (red curves), (ii) adiabatic transition (blue curves), and (iii) the combination of isothermal compression and diabatic transition (green curves). The processes in the $B$--$M$ diagrams evolve in a clockwise fashion over time (refrigerator), whereas the processes in the $A$--$D$ diagrams evolve in a counter-clockwise fashion over time (heat engine). The areas in each $B$--$M$ and $A$--$D$ diagram are identical; thus, the work done for the system is zero (see Table \ref{TableTwo}).}
\end{figure}

To demonstrate the roles of the isothermal and adiabatic transition processes, we conducted numerical simulations for a spin system expressed as
\begin{equation}
\label{MainSystemHamiltonian}
\hat{H}_A ( t ) = - B( t ) \hat{\sigma}_z ,
\end{equation}
where $B(t)$ is the IDF, and $\hat{\sigma}_{\alpha}$ ($\alpha = x, y,$ and $z$) are the Pauli matrices. In the case of nuclear magnetic resonance spectroscopy, $B(t)$ corresponds to the longitudinal magnetic field. The time-dependent SB interaction is controlled by the ATF [$A(t)$]. 

To construct the HEOM presented in Eq.~\eqref{ModelHEOM}, we assume the Drude spectral distribution function given by
\begin{equation}
J ( \omega) =\frac{\hbar }{\pi} \frac{\gamma^2 \omega}{\gamma^2 + \omega^2},
\end{equation}
where $\gamma$ is the inverse noise correlation time of the bath, and we set $\hat V= \hat \sigma_x$. We fix the inverse of the noise correlation time to $\gamma=1.0$ and use this as the frequency unit for the system. 
Then, we employ a Pad{\'e} spectral decomposition scheme to obtain the expansion coefficients of the noise correlation functions. 

Next, we consider three cases: (a) the high-temperature case ($\beta \hbar = 0.2$); (b) the intermediate-temperature case ($\beta \hbar = 1.0$); and (c) the low-temperature case ($\beta \hbar = 5.0$). We then choose the truncation number of hierarchy, which represents the depth of the HEOM computation, as $N = 6$. We set the maximum number of hierarchy levels to $K = 4$ for $\beta \hbar = 0.2$ and $1.0$ and $K = 7$ for $\beta \hbar = 5.0$. Starting from a temporal initial state, we integrate Eq.~\eqref{ModelHEOM} until the cycle of the simulation reaches the steady state with the time-dependent functions $B(t)$ and $A(t)$. We set the period of one stroke as $\tau = 10\,000$ so that the motion of the system is quasi-static. We use the fourth-order Runge--Kutta method with a time step of $\delta t = 1.0 \times 10^{-2}$. 

We conducted the simulation for a three-stroke engine (a Kelvin-Planck cycle) consisting of (i) isothermal expansion, (ii) isothermal--adiabatic transition, and (iii) the combination of isothermal compression and adiabatic--isothermal transition [described by $B(t)$ and $A(t)$ in Table \ref{Cycelarameter} with amplitudes of $B_0= 0.5$ and $A_0 = 1.0$, respectively]. To elucidate the characteristics of the cyclic process, we constructed thermodynamic work diagrams for external forces and their conjugate variables as the $B$--$M$ and $A$--$D$ diagrams analogous to the $P$--$V$ diagram. Figure \ref{BMADDiagram} depicts the $B$--$M$ (left) and $A$--$D$ (right) diagrams for different inverse temperatures. The processes in the $B$--$M$ diagrams evolve in a clockwise fashion over time, whereas those in the $A$--$D$ diagrams evolve in a counter-clockwise manner. In comparison with the $P$--$V$ diagram for an ideal gas, the rotational directions in the $B$--$M$ and $A$--$D$ diagrams are opposite because the ideal gas is described by $dU=TdS -PdV$, whereas here we have $dU=T dS +BdM+ AdD$.The area enclosed by the curves corresponds to the work, but a counterclockwise cycle represents positive work, which is also opposite to the $P$--$V$ case.

\renewcommand{\arraystretch}{1.3}
\begin{table}[h]
\caption{\label{TableTwo} System, interaction, and total quasi-static works per cycle ( $W_A^{qst}$, $W_I^{qst}$, and $W^{qst}_{tot}
=W_A^{qst} + W_I^{qst}$, respectively) at different inverse temperatures $\beta \hbar$. } 
\centering
\begin{tabular}{c|cc|c}
\hline
$\beta \hbar$ & $W_A^{qst}$ & $W_I^{qst}$ & $W_{tot}^{qst}$
\\
\hline
\hline
$\quad 0.2 \quad$ & $1.34 \times 10^{-3}$ & $-1.33 \times 10^{-3}$ & $1.04 \times 10^{-5}$
\\
$\quad 1.0 \quad$ & $1.71 \times 10^{-2}$ & $-1.71 \times 10^{-2}$ & $1.28 \times 10^{-5}$
\\
$\quad 5.0 \quad$ & $2.68 \times 10^{-2}$ & $-2.69 \times 10^{-2}$ & $-1.19 \times 10^{-4}$
\\
\hline
\end{tabular}
\end{table}
\renewcommand{\arraystretch}{1.0}

In this model, $B(t)$ represent the excitation energy of the spin. Then, as $B(t)$ increases, the spin is aligned with the ground state, so the magnetization $M( \beta; t)$ increases.
Since the SB interaction with $\hat V =\hat \sigma_x$ excites the spins, as $A(t)$ increases, $M( \beta; t)$ decreases, even if $B(t)$ does not change, as shown by the blue horizontal line in the $B$--$M$ diagram. Similarly, an increase in $B(t)$ suppresses the spin excitation effect of $A(t)$, so $D( \beta; t)$ decreases even if $A(t)$ does not change, as shown by the red horizontal line in the $A$--$D$ diagram.

Here, the areas surrounded by the counter-clockwise curves are positive work (a heat engine), whereas those surrounded by the clockwise curves are negative work (a refrigerator). This indicates that the IDF and ATF described in Table~\ref{Cycelarameter} drive the system as a refrigerator and heat engine, respectively. The size of each area is determined mainly by the adiabatic transition (the horizontal blue line) in the case of the $B$--$M$ diagrams and by the isothermal transition (the horizontal red line) in the case of the $A$--$D$ diagrams. As the temperature decreases, the area becomes larger because the efficiency of the energy change under a given external force improves as the internal energy decreases. Thus, we find that the areas in the $B$--$M$ and $A$--$D$ diagrams are identical at each temperature, and the work done for the system is zero within the numerical accuracy (see Table~\ref{TableTwo}) because we cannot subtract energy from the heat machine with a single heat bath. This result demonstrates the Kelvin--Planck statement (or heat engine statement) of the second law of thermodynamics\cite{Oono2017} in open quantum dynamics theory: it is impossible for any substance to derive mechanical effects from a single heat source.

\section{CONCLUSIONS}
\label{conclusion}
We investigated the thermodynamic properties of quantum dissipative systems based on the SB model by identifying the quasi-static work as the qHE. The key to investigating the non-equilibrium thermodynamic problem is the HEOM formalism, which enables the evaluation of the internal energies of not only the system, but also the bath and the SB interaction, even under low-temperature, non-Markovian, and nonperturbative conditions, where the quantum effects become important. While the qHE was originally developed for an isothermal process, we extended it to treat an adiabatic transition process in a unified manner. 

As a demonstration, we numerically simulated a three-stroke heat machine consisting of an isothermal process, an adiabatic process, and their combination. To analyze the results, the work diagrams with external fields and their conjugated variables were used. The results are consistent with the Kelvin--Planck statement of the second law of thermodynamics. This indicates that the thermodynamic rule is broken if the work of adiabatic wall manipulation described by the ATF is not taken into account, suggesting its importance.
In the case of Markov limits where the $\gamma$ is large and the heat bath is hot, or in the case of perturbative approximations where the interaction is weak, the effects of the ATF are expected to be less pronounced. 

Although our demonstration was restricted to a simple model, this approach can be applied to investigate a variety of heat engines and refrigerators that consist of the isothermal and adiabatic processes. Moreover, while we analyzed only a quasi-static case, the present formalism can be applied to non-equilibrium situations by regarding the work as the non-equilibrium free energy. Numerically rigorous HEOM experiments on such systems can be a versatile means to formulate and verify quantum thermodynamics far from equilibrium. In the future, we plan to extend the present research to a study of the quantum Carnot cycle.

\begin{acknowledgments}
Y.T. is supported by JSPS KAKENHI Grant No. B 21H01884.
S.K. acknowledges a fellowship supported by JST, the establishment of university fellowships towards the creation of science technology innovation, Grant Number JPMJFS2123.
\end{acknowledgments}

\section*{AUTHOR DECLARATIONS}
\subsection*{Conflict of Interest}
The authors have no conflicts to disclose.

\section*{Data availability}
The data that support the findings of this study are available from the corresponding author upon reasonable request.

\appendix
\section{The minimum work principle [Eq.\eqref{Ineq}]}
\label{AppQHE}

\begin{figure}[h]
\centering
\includegraphics[width=6cm]{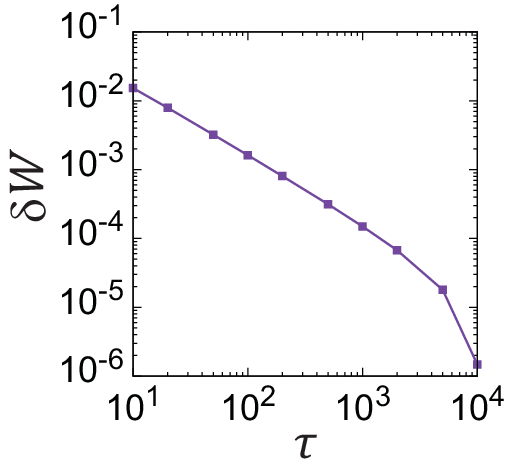}
\caption{\label{FigqHE}$\delta W(\tau) \equiv W_{tot}(\tau) - \Delta F_{tot}(\tau)$ plotted as a function of time duration $\tau$. For any $\tau$,  $\delta W(\tau) $ is positive, indicating that the inequality Eq.~\eqref{Ineq} holds.} 
\end{figure}

\begin{table}[h]
\caption{\label{AppQHETable} The IDF [$B ( t )$]  and ATF [$A ( t )$] as functions of $t$. Here, $\tau$ is the time duration, and the process approaches a quasi-static state when $\tau$ becomes large. }
\centering
\begin{tabular}{cccc}
\hline
time & $B ( t ) $ & $A ( t ) $
\\
\hline
\hline
$t < 0$ & $1.0$ & $1.0$
\\
$\quad 0 \leq t < \tau \quad$ & $\quad 1.0 + t / \tau \quad$ & $\quad 1.0 - 0.3 \; t / \tau \quad$
\\
$\tau \leq t$ & $2.0$ & $0.7$
\\
\hline
\end{tabular}
\end{table}
For an isolated quantum system described by the Hamiltonian ${\hat H}_{0}$ and the perturbed Hamiltonian ${\hat H}_{\delta t}= \hat H_{0} + \delta \hat H $, we have an identity that leads to a quantum version of the Jarzynski equality expressed as\cite{tasaki2000, Crooks_2008,YukawaJPSJ00,HanggiPRE2007}
\begin{align}
\left\langle e^{- \beta \hat{H}_0^{( H )} ( \delta t )} e^{ \beta \hat{H}_0} \right\rangle
 = \frac{Z_{\delta t}}{Z_0},
\label{eq:Hanggi}
\end{align}
where $\langle ...  \rangle$ implies an average of appropriate sampling states and $Z_{\delta t}$ and $Z_{0}$ are the partition functions for $\hat H_{\delta t}$ and $\hat H_0$ at the inverse temperature $\beta$ and ${\hat H}_0^{(H)}(\delta t)$ is the Heisenberg operator of ${\hat H}_0$ whose time dependence is described by ${\hat H}_{\delta t}$. 

The orthonormal basis set of ${\hat H}_0$ is expressed as $\lbrace |i \rangle \rbrace$ of ${\hat H}_0$. Then we have inequality expressed as\cite{Oono2017}
\begin{align}
\langle i | e^{- \beta \hat{H}} | i \rangle \geq e^{\langle i | \hat{H} | i \rangle},
\label{eq:ProtoPeierls}
\end{align}
for any Hermitian operator $\hat{H}$.
This leads to 
\begin{align}
\left\langle e^{- \beta \hat{H}_0^{( H )} ( \delta t )} e^{ \beta \hat{H}_0} \right\rangle
\geq \sum_i p_i e^{- \beta ( \langle i | \hat{H}_0^{( H )} ( \delta t ) | i \rangle - E_i )} ,
\end{align}
where $E_i = \langle i | \hat{H}_0 | i \rangle$ and $p_i = e^{- \beta E_i} / Z_0$.
Using Jensen's inequality
$\langle \exp [ - \beta \hat X ]  \rangle \ge \exp[-\beta \langle  \hat X    \rangle ]$, we have\cite{Lenard1978,tasaki2000,Hanggi2011FT}
\begin{align}
- \beta \left(  \langle {\hat H}_0^{(H)} (\delta t) \rangle -  \langle {\hat H}_0 \rangle \right) 
\leq \ln Z_{\delta t} - \ln Z_0.
\label{eq:Tasaki}
\end{align}
Because our SB system is also an isolated system, the above inequality holds for ${\hat H}_{tot}(t)$ with the heat bath Eq. \eqref{Bath} consisting of $N$ harmonic oscillators.  Thus we obtain 
\begin{align}
W_{tot} (\beta ; t) \ge \Delta F_{tot} (\beta; t),
\label{Ineq2} 
\end{align}
which is the minimum work principle given by Eq.\eqref{Ineq}. Finally, taking the limit of $N \rightarrow \infty$, we obtain the description for the case of open quantum dynamics systems.
In general, the above inequalities may not hold for reduced systems, because both work and free energy are evaluated from a reduced system rather than from an isolated whole system: using the HEOM formalism, the above inequality can be applied because contributions from the bath part of work and free energy can also be evaluated.

In Eq. \eqref{Ineq2}, the equal sign holds if the process is reversible (the minimum work principle).  The quantity $\delta W(\beta ; t)= W_{tot} (\beta ; t) - \Delta F_{tot}(\beta ; t )$ produced in the irreversible case is the work discarded as dissipative heat; in a quasi-static process, the energy loss due to dissipation is suppressed and $ W_{tot} (\beta ; t)$ approaches  $\Delta F_{tot}(\beta ; t )$.

Although Eq. \eqref{Ineq2} for quantum finite systems has been proved, the mathematical proof for infinite systems is not complete at this time. Here we conduct numerical calculations using the HEOM, which are derived by reducing a heat bath with infinite degrees of freedom, and verified the results within the error range.

For the SB Hamiltonian, Eq.~\eqref{Ineq} has been examined numerically in the isothermal case.\cite{ST20JCP} Here, we extend it to the case of a mixture of isothermal and adiabatic transitions using the system discussed in Sec.~\ref{numericalresults}. To conduct the simulation, we set the IDF and ATF as in Table \ref{AppQHETable}. The non-equilibrium work $W_{tot} ( \tau )$ is evaluated from Eq.~\eqref{ModelHEOM} for various values of $\tau$ and fixed parameter values ($\gamma = 1.0, \beta \hbar = 1.0, N = 8, K = 4$, and $\delta t = 0.01$). Meanwhile, the equilibrium free energy $\Delta F_{tot} (\tau)$ is evaluated from the partition function obtained from the imaginary HEOM with $N = 6$ and $K = 10$ and the imaginary time step $\delta t = 0.0001.$\cite{T14JCP}   
We plot $\delta W(\tau)\equiv W_{tot} (\tau) - \Delta F_{tot}(\tau )$ for different time scale of driving fields $\tau$ in Fig. \ref{FigqHE}. As shown in this figure, $\delta F$ is always positive.

\section{DERIVATION OF Eq.~\eqref{expectHB}}
\label{AppExpectHB}
The time differentiation of the expectation value $\mathrm{tr}_{tot} \{ \hat{H}_{tot} ( t ) \hat{\rho}_{tot} ( t ) \}$ is expressed as
\begin{equation}
\label{AppUB}
\frac{\partial U_B ( t )}{\partial t} = W_{tot} ( t ) - \frac{\partial U_A ( t )}{\partial t}
- \frac{\partial U_I ( t )}{\partial t},
\end{equation}
where $W_{tot} ( t )$ is defined in Eq.~\eqref{workWtot}. Applying Eqs.~\eqref{expectHA} and \eqref{expectHI} in Eq.~\eqref{AppUB}, we can rewrite Eq.~\eqref{AppUB} as
\begin{equation}
\label{AppUBTwo}
\frac{\partial U_B ( t )}{\partial t} = - \mathrm{tr}_A \left\{ \hat{H}_A ( t ) 
\frac{\partial \hat{\rho}_{\vec{0}} ( t )}{\partial t} \right\}
- A ( t ) \sum_{l = 0}^K \mathrm{tr}_A \left\{ \hat{V} 
\frac{\partial \hat{\rho}_{\vec{e}_l} ( t )}{\partial t} \right\}.
\end{equation}
The second term on the right-hand side of the above equation can be evaluated from Eq.~\eqref{ModelHEOM}. Thus, we obtain Eq.~\eqref{expectHB}.

%%%%%%%%%%%%

%%%%%%%%%%%%
\bibliography{tanimura_publist,referencesqHE}

\begin{thebibliography}{83}%
\makeatletter
\providecommand \@ifxundefined [1]{%
 \@ifx{#1\undefined}
}%
\providecommand \@ifnum [1]{%
 \ifnum #1\expandafter \@firstoftwo
 \else \expandafter \@secondoftwo
 \fi
}%
\providecommand \@ifx [1]{%
 \ifx #1\expandafter \@firstoftwo
 \else \expandafter \@secondoftwo
 \fi
}%
\providecommand \natexlab [1]{#1}%
\providecommand \enquote  [1]{``#1''}%
\providecommand \bibnamefont  [1]{#1}%
\providecommand \bibfnamefont [1]{#1}%
\providecommand \citenamefont [1]{#1}%
\providecommand \href@noop [0]{\@secondoftwo}%
\providecommand \href [0]{\begingroup \@sanitize@url \@href}%
\providecommand \@href[1]{\@@startlink{#1}\@@href}%
\providecommand \@@href[1]{\endgroup#1\@@endlink}%
\providecommand \@sanitize@url [0]{\catcode `\\12\catcode `\$12\catcode
  `\&12\catcode `\#12\catcode `\^12\catcode `\_12\catcode `\%12\relax}%
\providecommand \@@startlink[1]{}%
\providecommand \@@endlink[0]{}%
\providecommand \url  [0]{\begingroup\@sanitize@url \@url }%
\providecommand \@url [1]{\endgroup\@href {#1}{\urlprefix }}%
\providecommand \urlprefix  [0]{URL }%
\providecommand \Eprint [0]{\href }%
\providecommand \doibase [0]{http://dx.doi.org/}%
\providecommand \selectlanguage [0]{\@gobble}%
\providecommand \bibinfo  [0]{\@secondoftwo}%
\providecommand \bibfield  [0]{\@secondoftwo}%
\providecommand \translation [1]{[#1]}%
\providecommand \BibitemOpen [0]{}%
\providecommand \bibitemStop [0]{}%
\providecommand \bibitemNoStop [0]{.\EOS\space}%
\providecommand \EOS [0]{\spacefactor3000\relax}%
\providecommand \BibitemShut  [1]{\csname bibitem#1\endcsname}%
\let\auto@bib@innerbib\@empty
%</preamble>
\bibitem [{\citenamefont {Gibbs}(1876)}]{Gibbs1876}%
  \BibitemOpen
  \bibfield  {author} {\bibinfo {author} {\bibfnamefont {J.~W.}\ \bibnamefont
  {Gibbs}},\ }\bibfield  {title} {\enquote {\bibinfo {title} {On the
  equilibrium of heterogeneous substances},}\ }\href@noop {} {\bibfield
  {journal} {\bibinfo  {journal} {Transactions of the Connecticut Academy of
  Arts and Sciences}\ }\textbf {\bibinfo {volume} {3}},\ \bibinfo {pages}
  {108--248} (\bibinfo {year} {1876})}\BibitemShut {NoStop}%
\bibitem [{\citenamefont {Gibbs}(1878)}]{Gibbs1878}%
  \BibitemOpen
  \bibfield  {author} {\bibinfo {author} {\bibfnamefont {J.~W.}\ \bibnamefont
  {Gibbs}},\ }\bibfield  {title} {\enquote {\bibinfo {title} {On the
  equilibrium of heterogeneous substances},}\ }\href@noop {} {\bibfield
  {journal} {\bibinfo  {journal} {Transactions of the Connecticut Academy of
  Arts and Sciences}\ }\textbf {\bibinfo {volume} {3}},\ \bibinfo {pages}
  {343--524} (\bibinfo {year} {1878})}\BibitemShut {NoStop}%
\bibitem [{\citenamefont {Helmholtz}(1882)}]{Helmholtz1882}%
  \BibitemOpen
  \bibfield  {author} {\bibinfo {author} {\bibfnamefont {H.}~\bibnamefont
  {Helmholtz}},\ }\href@noop {} {\emph {\bibinfo {title} {Wissenschaftliche
  abhandlungen}}}\ (\bibinfo  {publisher} {Leipzig, J. A. Barth},\ \bibinfo
  {year} {1882})\BibitemShut {NoStop}%
\bibitem [{\citenamefont {Oono}(2017)}]{Oono2017}%
  \BibitemOpen
  \bibfield  {author} {\bibinfo {author} {\bibfnamefont {Y.}~\bibnamefont
  {Oono}},\ }\href {\doibase 10.1017/9781316650394} {\emph {\bibinfo {title}
  {Perspectives on Statistical Thermodynamics}}}\ (\bibinfo  {publisher}
  {Cambridge University Press},\ \bibinfo {year} {2017})\BibitemShut {NoStop}%
\bibitem [{\citenamefont {Jarzynski}(1997)}]{JarzynskiPRL1997}%
  \BibitemOpen
  \bibfield  {author} {\bibinfo {author} {\bibfnamefont {C.}~\bibnamefont
  {Jarzynski}},\ }\bibfield  {title} {\enquote {\bibinfo {title}
  {Nonequilibrium equality for free energy differences},}\ }\href {\doibase
  10.1103/PhysRevLett.78.2690} {\bibfield  {journal} {\bibinfo  {journal}
  {Phys. Rev. Lett.}\ }\textbf {\bibinfo {volume} {78}},\ \bibinfo {pages}
  {2690--2693} (\bibinfo {year} {1997})}\BibitemShut {NoStop}%
\bibitem [{\citenamefont {Jarzynski}(2004)}]{Jarzynski2004}%
  \BibitemOpen
  \bibfield  {author} {\bibinfo {author} {\bibfnamefont {C.}~\bibnamefont
  {Jarzynski}},\ }\bibfield  {title} {\enquote {\bibinfo {title}
  {Nonequilibrium work theorem for a system strongly coupled to a thermal
  environment},}\ }\href {\doibase 10.1088/1742-5468/2004/09/p09005} {\bibfield
   {journal} {\bibinfo  {journal} {Journal of Statistical Mechanics: Theory and
  Experiment}\ }\textbf {\bibinfo {volume} {2004}},\ \bibinfo {pages} {P09005}
  (\bibinfo {year} {2004})}\BibitemShut {NoStop}%
\bibitem [{\citenamefont {Jarzynski}(2011)}]{JarzynskiAnnu11}%
  \BibitemOpen
  \bibfield  {author} {\bibinfo {author} {\bibfnamefont {C.}~\bibnamefont
  {Jarzynski}},\ }\bibfield  {title} {\enquote {\bibinfo {title} {Equalities
  and inequalities: Irreversibility and the second law of thermodynamics at the
  nanoscale},}\ }\href {\doibase 10.1146/annurev-conmatphys-062910-140506}
  {\bibfield  {journal} {\bibinfo  {journal} {Annual Review of Condensed Matter
  Physics}\ }\textbf {\bibinfo {volume} {2}},\ \bibinfo {pages} {329--351}
  (\bibinfo {year} {2011})}\BibitemShut {NoStop}%
\bibitem [{\citenamefont {Tasaki}(2000)}]{tasaki2000}%
  \BibitemOpen
  \bibfield  {author} {\bibinfo {author} {\bibfnamefont {H.}~\bibnamefont
  {Tasaki}},\ }\bibfield  {title} {\enquote {\bibinfo {title} {Jarzynski
  relations for quantum systems and some applications},}\ }\href@noop {} {\
  (\bibinfo {year} {2000})},\ \Eprint {http://arxiv.org/abs/cond-mat/0009244}
  {arXiv:cond-mat/0009244 [cond-mat.stat-mech]} \BibitemShut {NoStop}%
\bibitem [{\citenamefont {Kurchan}(2001)}]{kurchan2001quantum}%
  \BibitemOpen
  \bibfield  {author} {\bibinfo {author} {\bibfnamefont {J.}~\bibnamefont
  {Kurchan}},\ }\href@noop {} {\enquote {\bibinfo {title} {A quantum
  fluctuation theorem},}\ } (\bibinfo {year} {2001}),\ \Eprint
  {http://arxiv.org/abs/cond-mat/0007360} {arXiv:cond-mat/0007360
  [cond-mat.stat-mech]} \BibitemShut {NoStop}%
\bibitem [{\citenamefont {Crooks}(2008)}]{Crooks_2008}%
  \BibitemOpen
  \bibfield  {author} {\bibinfo {author} {\bibfnamefont {G.~E.}\ \bibnamefont
  {Crooks}},\ }\bibfield  {title} {\enquote {\bibinfo {title} {On the jarzynski
  relation for dissipative quantum dynamics},}\ }\href {\doibase
  10.1088/1742-5468/2008/10/p10023} {\bibfield  {journal} {\bibinfo  {journal}
  {Journal of Statistical Mechanics: Theory and Experiment}\ }\textbf {\bibinfo
  {volume} {2008}},\ \bibinfo {pages} {P10023} (\bibinfo {year}
  {2008})}\BibitemShut {NoStop}%
\bibitem [{\citenamefont {Yukawa}(2000)}]{YukawaJPSJ00}%
  \BibitemOpen
  \bibfield  {author} {\bibinfo {author} {\bibfnamefont {S.}~\bibnamefont
  {Yukawa}},\ }\bibfield  {title} {\enquote {\bibinfo {title} {A quantum
  analogue of the jarzynski equality},}\ }\href {\doibase 10.1143/JPSJ.69.2367}
  {\bibfield  {journal} {\bibinfo  {journal} {Journal of the Physical Society
  of Japan}\ }\textbf {\bibinfo {volume} {69}},\ \bibinfo {pages} {2367--2370}
  (\bibinfo {year} {2000})}\BibitemShut {NoStop}%
\bibitem [{\citenamefont {Esposito}\ and\ \citenamefont
  {Mukamel}(2006)}]{Mukamel2006FT}%
  \BibitemOpen
  \bibfield  {author} {\bibinfo {author} {\bibfnamefont {M.}~\bibnamefont
  {Esposito}}\ and\ \bibinfo {author} {\bibfnamefont {S.}~\bibnamefont
  {Mukamel}},\ }\bibfield  {title} {\enquote {\bibinfo {title} {Fluctuation
  theorems for quantum master equations},}\ }\href {\doibase
  10.1103/PhysRevE.73.046129} {\bibfield  {journal} {\bibinfo  {journal} {Phys.
  Rev. E}\ }\textbf {\bibinfo {volume} {73}},\ \bibinfo {pages} {046129}
  (\bibinfo {year} {2006})}\BibitemShut {NoStop}%
\bibitem [{\citenamefont {Esposito}, \citenamefont {Harbola},\ and\
  \citenamefont {Mukamel}(2009)}]{Mukamel2019FT}%
  \BibitemOpen
  \bibfield  {author} {\bibinfo {author} {\bibfnamefont {M.}~\bibnamefont
  {Esposito}}, \bibinfo {author} {\bibfnamefont {U.}~\bibnamefont {Harbola}}, \
  and\ \bibinfo {author} {\bibfnamefont {S.}~\bibnamefont {Mukamel}},\
  }\bibfield  {title} {\enquote {\bibinfo {title} {Nonequilibrium fluctuations,
  fluctuation theorems, and counting statistics in quantum systems},}\ }\href
  {\doibase 10.1103/RevModPhys.81.1665} {\bibfield  {journal} {\bibinfo
  {journal} {Rev. Mod. Phys.}\ }\textbf {\bibinfo {volume} {81}},\ \bibinfo
  {pages} {1665--1702} (\bibinfo {year} {2009})}\BibitemShut {NoStop}%
\bibitem [{\citenamefont {Talkner}, \citenamefont {Lutz},\ and\ \citenamefont
  {H\"anggi}(2007)}]{HanggiPRE2007}%
  \BibitemOpen
  \bibfield  {author} {\bibinfo {author} {\bibfnamefont {P.}~\bibnamefont
  {Talkner}}, \bibinfo {author} {\bibfnamefont {E.}~\bibnamefont {Lutz}}, \
  and\ \bibinfo {author} {\bibfnamefont {P.}~\bibnamefont {H\"anggi}},\
  }\bibfield  {title} {\enquote {\bibinfo {title} {Fluctuation theorems: Work
  is not an observable},}\ }\href {\doibase 10.1103/PhysRevE.75.050102}
  {\bibfield  {journal} {\bibinfo  {journal} {Phys. Rev. E}\ }\textbf {\bibinfo
  {volume} {75}},\ \bibinfo {pages} {050102} (\bibinfo {year}
  {2007})}\BibitemShut {NoStop}%
\bibitem [{\citenamefont {Campisi}, \citenamefont {Talkner},\ and\
  \citenamefont {H\"anggi}(2009)}]{Hanggi2009FT}%
  \BibitemOpen
  \bibfield  {author} {\bibinfo {author} {\bibfnamefont {M.}~\bibnamefont
  {Campisi}}, \bibinfo {author} {\bibfnamefont {P.}~\bibnamefont {Talkner}}, \
  and\ \bibinfo {author} {\bibfnamefont {P.}~\bibnamefont {H\"anggi}},\
  }\bibfield  {title} {\enquote {\bibinfo {title} {Fluctuation theorem for
  arbitrary open quantum systems},}\ }\href {\doibase
  10.1103/PhysRevLett.102.210401} {\bibfield  {journal} {\bibinfo  {journal}
  {Phys. Rev. Lett.}\ }\textbf {\bibinfo {volume} {102}},\ \bibinfo {pages}
  {210401} (\bibinfo {year} {2009})}\BibitemShut {NoStop}%
\bibitem [{\citenamefont {Campisi}, \citenamefont {H\"anggi},\ and\
  \citenamefont {Talkner}(2011)}]{Hanggi2011FT}%
  \BibitemOpen
  \bibfield  {author} {\bibinfo {author} {\bibfnamefont {M.}~\bibnamefont
  {Campisi}}, \bibinfo {author} {\bibfnamefont {P.}~\bibnamefont {H\"anggi}}, \
  and\ \bibinfo {author} {\bibfnamefont {P.}~\bibnamefont {Talkner}},\
  }\bibfield  {title} {\enquote {\bibinfo {title} {Colloquium: Quantum
  fluctuation relations: Foundations and applications},}\ }\href {\doibase
  10.1103/RevModPhys.83.771} {\bibfield  {journal} {\bibinfo  {journal} {Rev.
  Mod. Phys.}\ }\textbf {\bibinfo {volume} {83}},\ \bibinfo {pages} {771--791}
  (\bibinfo {year} {2011})}\BibitemShut {NoStop}%
\bibitem [{\citenamefont {Talkner}\ and\ \citenamefont
  {H\"anggi}(2020)}]{Hanggi2020}%
  \BibitemOpen
  \bibfield  {author} {\bibinfo {author} {\bibfnamefont {P.}~\bibnamefont
  {Talkner}}\ and\ \bibinfo {author} {\bibfnamefont {P.}~\bibnamefont
  {H\"anggi}},\ }\bibfield  {title} {\enquote {\bibinfo {title} {Colloquium:
  Statistical mechanics and thermodynamics at strong coupling: Quantum and
  classical},}\ }\href {\doibase 10.1103/RevModPhys.92.041002} {\bibfield
  {journal} {\bibinfo  {journal} {Rev. Mod. Phys.}\ }\textbf {\bibinfo {volume}
  {92}},\ \bibinfo {pages} {041002} (\bibinfo {year} {2020})}\BibitemShut
  {NoStop}%
\bibitem [{\citenamefont {Crooks}(1999)}]{Crooks1999}%
  \BibitemOpen
  \bibfield  {author} {\bibinfo {author} {\bibfnamefont {G.~E.}\ \bibnamefont
  {Crooks}},\ }\bibfield  {title} {\enquote {\bibinfo {title} {Entropy
  production fluctuation theorem and the nonequilibrium work relation for free
  energy differences},}\ }\href {\doibase 10.1103/PhysRevE.60.2721} {\bibfield
  {journal} {\bibinfo  {journal} {Phys. Rev. E}\ }\textbf {\bibinfo {volume}
  {60}},\ \bibinfo {pages} {2721--2726} (\bibinfo {year} {1999})}\BibitemShut
  {NoStop}%
\bibitem [{\citenamefont {Segal}(2006)}]{Segal2006}%
  \BibitemOpen
  \bibfield  {author} {\bibinfo {author} {\bibfnamefont {D.}~\bibnamefont
  {Segal}},\ }\bibfield  {title} {\enquote {\bibinfo {title} {Heat flow in
  nonlinear molecular junctions: Master equation analysis},}\ }\href {\doibase
  10.1103/PhysRevB.73.205415} {\bibfield  {journal} {\bibinfo  {journal} {Phys.
  Rev. B}\ }\textbf {\bibinfo {volume} {73}},\ \bibinfo {pages} {205415}
  (\bibinfo {year} {2006})}\BibitemShut {NoStop}%
\bibitem [{\citenamefont {Velizhanin}, \citenamefont {Wang},\ and\
  \citenamefont {Thoss}(2008)}]{Thoss2008}%
  \BibitemOpen
  \bibfield  {author} {\bibinfo {author} {\bibfnamefont {K.~A.}\ \bibnamefont
  {Velizhanin}}, \bibinfo {author} {\bibfnamefont {H.}~\bibnamefont {Wang}}, \
  and\ \bibinfo {author} {\bibfnamefont {M.}~\bibnamefont {Thoss}},\ }\bibfield
   {title} {\enquote {\bibinfo {title} {Heat transport through model molecular
  junctions: A multilayer multiconfiguration time-dependent hartree
  approach},}\ }\href {\doibase https://doi.org/10.1016/j.cplett.2008.05.065}
  {\bibfield  {journal} {\bibinfo  {journal} {Chemical Physics Letters}\
  }\textbf {\bibinfo {volume} {460}},\ \bibinfo {pages} {325--330} (\bibinfo
  {year} {2008})}\BibitemShut {NoStop}%
\bibitem [{\citenamefont {Dubi}\ and\ \citenamefont
  {Di~Ventra}(2011)}]{YonatanRMP2011}%
  \BibitemOpen
  \bibfield  {author} {\bibinfo {author} {\bibfnamefont {Y.}~\bibnamefont
  {Dubi}}\ and\ \bibinfo {author} {\bibfnamefont {M.}~\bibnamefont
  {Di~Ventra}},\ }\bibfield  {title} {\enquote {\bibinfo {title} {Colloquium:
  Heat flow and thermoelectricity in atomic and molecular junctions},}\ }\href
  {\doibase 10.1103/RevModPhys.83.131} {\bibfield  {journal} {\bibinfo
  {journal} {Rev. Mod. Phys.}\ }\textbf {\bibinfo {volume} {83}},\ \bibinfo
  {pages} {131--155} (\bibinfo {year} {2011})}\BibitemShut {NoStop}%
\bibitem [{\citenamefont {Saito}\ and\ \citenamefont
  {Kato}(2013)}]{SaitoPRL2013}%
  \BibitemOpen
  \bibfield  {author} {\bibinfo {author} {\bibfnamefont {K.}~\bibnamefont
  {Saito}}\ and\ \bibinfo {author} {\bibfnamefont {T.}~\bibnamefont {Kato}},\
  }\bibfield  {title} {\enquote {\bibinfo {title} {Kondo signature in heat
  transfer via a local two-state system},}\ }\href {\doibase
  10.1103/PhysRevLett.111.214301} {\bibfield  {journal} {\bibinfo  {journal}
  {Phys. Rev. Lett.}\ }\textbf {\bibinfo {volume} {111}},\ \bibinfo {pages}
  {214301} (\bibinfo {year} {2013})}\BibitemShut {NoStop}%
\bibitem [{\citenamefont {Wang}, \citenamefont {Ren},\ and\ \citenamefont
  {Cao}(2015)}]{2015JCao}%
  \BibitemOpen
  \bibfield  {author} {\bibinfo {author} {\bibfnamefont {C.}~\bibnamefont
  {Wang}}, \bibinfo {author} {\bibfnamefont {J.}~\bibnamefont {Ren}}, \ and\
  \bibinfo {author} {\bibfnamefont {J.}~\bibnamefont {Cao}},\ }\bibfield
  {title} {\enquote {\bibinfo {title} {Nonequilibrium energy transfer at
  nanoscale: A unified theory from weak to strong coupling},}\ }\href {\doibase
  10.1038/srep11787} {\bibfield  {journal} {\bibinfo  {journal} {Scientific
  Reports}\ }\textbf {\bibinfo {volume} {5}} (\bibinfo {year} {2015}),\
  10.1038/srep11787}\BibitemShut {NoStop}%
\bibitem [{\citenamefont {de~Vega}\ and\ \citenamefont
  {Alonso}(2017)}]{Alonso2017}%
  \BibitemOpen
  \bibfield  {author} {\bibinfo {author} {\bibfnamefont {I.}~\bibnamefont
  {de~Vega}}\ and\ \bibinfo {author} {\bibfnamefont {D.}~\bibnamefont
  {Alonso}},\ }\bibfield  {title} {\enquote {\bibinfo {title} {Dynamics of
  non-markovian open quantum systems},}\ }\href {\doibase
  10.1103/RevModPhys.89.015001} {\bibfield  {journal} {\bibinfo  {journal}
  {Rev. Mod. Phys.}\ }\textbf {\bibinfo {volume} {89}},\ \bibinfo {pages}
  {015001} (\bibinfo {year} {2017})}\BibitemShut {NoStop}%
\bibitem [{\citenamefont {Whitney}(2018)}]{WhitneyPhysRevB2018}%
  \BibitemOpen
  \bibfield  {author} {\bibinfo {author} {\bibfnamefont {R.~S.}\ \bibnamefont
  {Whitney}},\ }\bibfield  {title} {\enquote {\bibinfo {title} {Non-markovian
  quantum thermodynamics: Laws and fluctuation theorems},}\ }\href {\doibase
  10.1103/PhysRevB.98.085415} {\bibfield  {journal} {\bibinfo  {journal} {Phys.
  Rev. B}\ }\textbf {\bibinfo {volume} {98}},\ \bibinfo {pages} {085415}
  (\bibinfo {year} {2018})}\BibitemShut {NoStop}%
\bibitem [{\citenamefont {Cohen}\ and\ \citenamefont
  {Galperin}(2020)}]{GalperinJCP2020}%
  \BibitemOpen
  \bibfield  {author} {\bibinfo {author} {\bibfnamefont {G.}~\bibnamefont
  {Cohen}}\ and\ \bibinfo {author} {\bibfnamefont {M.}~\bibnamefont
  {Galperin}},\ }\bibfield  {title} {\enquote {\bibinfo {title} {Green’s
  function methods for single molecule junctions},}\ }\href {\doibase
  10.1063/1.5145210} {\bibfield  {journal} {\bibinfo  {journal} {The Journal of
  Chemical Physics}\ }\textbf {\bibinfo {volume} {152}},\ \bibinfo {pages}
  {090901} (\bibinfo {year} {2020})},\ \Eprint
  {http://arxiv.org/abs/https://doi.org/10.1063/1.5145210}
  {https://doi.org/10.1063/1.5145210} \BibitemShut {NoStop}%
\bibitem [{\citenamefont {Ryu}\ \emph {et~al.}(2022)\citenamefont {Ryu},
  \citenamefont {López}, \citenamefont {Serra},\ and\ \citenamefont
  {Sánchez}}]{SanchezdNatureCarnot2022}%
  \BibitemOpen
  \bibfield  {author} {\bibinfo {author} {\bibfnamefont {S.}~\bibnamefont
  {Ryu}}, \bibinfo {author} {\bibfnamefont {R.}~\bibnamefont {López}},
  \bibinfo {author} {\bibfnamefont {L.}~\bibnamefont {Serra}}, \ and\ \bibinfo
  {author} {\bibfnamefont {D.}~\bibnamefont {Sánchez}},\ }\bibfield  {title}
  {\enquote {\bibinfo {title} {Beating carnot efficiency with periodically
  driven chiral conductorse},}\ }\href {\doibase 10.1038/s41467-022-30039-7}
  {\bibfield  {journal} {\bibinfo  {journal} {Nature Communications}\ }\textbf
  {\bibinfo {volume} {13}},\ \bibinfo {pages} {2512} (\bibinfo {year}
  {2022})}\BibitemShut {NoStop}%
\bibitem [{\citenamefont {Breuer}\ and\ \citenamefont
  {Petruccione}(2002)}]{Breuer2002}%
  \BibitemOpen
  \bibfield  {author} {\bibinfo {author} {\bibfnamefont {H.-P.}\ \bibnamefont
  {Breuer}}\ and\ \bibinfo {author} {\bibfnamefont {F.}~\bibnamefont
  {Petruccione}},\ }\href@noop {} {\emph {\bibinfo {title} {The theory of open
  quantum systems}}}\ (\bibinfo  {publisher} {Oxford:Oxford University Press},\
  \bibinfo {year} {2002})\ pp.\ \bibinfo {pages} {xxi + 625}\BibitemShut
  {NoStop}%
\bibitem [{\citenamefont {Weiss}(2012)}]{Weiss2012}%
  \BibitemOpen
  \bibfield  {author} {\bibinfo {author} {\bibfnamefont {U.}~\bibnamefont
  {Weiss}},\ }\href {\doibase 10.1142/8334} {\emph {\bibinfo {title} {Quantum
  Dissipative Systems}}},\ \bibinfo {edition} {4th}\ ed.\ (\bibinfo
  {publisher} {WORLD SCIENTIFIC},\ \bibinfo {year} {2012})\BibitemShut
  {NoStop}%
\bibitem [{\citenamefont {Binder}\ \emph {et~al.}(2018)\citenamefont {Binder},
  \citenamefont {Correa}, \citenamefont {C}, \citenamefont {Anders},\ and\
  \citenamefont {(Eds.)}}]{Binder2018}%
  \BibitemOpen
  \bibfield  {author} {\bibinfo {author} {\bibfnamefont {F.}~\bibnamefont
  {Binder}}, \bibinfo {author} {\bibfnamefont {L.~A.}\ \bibnamefont {Correa}},
  \bibinfo {author} {\bibfnamefont {G.}~\bibnamefont {C}}, \bibinfo {author}
  {\bibfnamefont {J.}~\bibnamefont {Anders}}, \ and\ \bibinfo {author}
  {\bibfnamefont {G.~A.}\ \bibnamefont {(Eds.)}},\ }\href {\doibase
  10.1007/978-3-319-99046-0} {\emph {\bibinfo {title} {Thermodynamics in the
  Quantum Regime}}}\ (\bibinfo  {publisher} {Springer International
  Publishing},\ \bibinfo {year} {2018})\BibitemShut {NoStop}%
\bibitem [{\citenamefont {Tanimura}(2006)}]{T06JPSJ}%
  \BibitemOpen
  \bibfield  {author} {\bibinfo {author} {\bibfnamefont {Y.}~\bibnamefont
  {Tanimura}},\ }\bibfield  {title} {\enquote {\bibinfo {title} {Stochastic
  liouville, langevin, fokker-planck, and master equation qpproaches to quantum
  dissipative systems},}\ }\href {\doibase 10.1143/JPSJ.75.082001} {\bibfield
  {journal} {\bibinfo  {journal} {Journal of the Physical Society of Japan}\
  }\textbf {\bibinfo {volume} {75}},\ \bibinfo {pages} {082001} (\bibinfo
  {year} {2006})}\BibitemShut {NoStop}%
\bibitem [{\citenamefont {Tanimura}(2014)}]{T14JCP}%
  \BibitemOpen
  \bibfield  {author} {\bibinfo {author} {\bibfnamefont {Y.}~\bibnamefont
  {Tanimura}},\ }\bibfield  {title} {\enquote {\bibinfo {title} {Reduced
  hierarchical equations of motion in real and imaginary time: Correlated
  initial states and thermodynamic quantities},}\ }\href {\doibase
  10.1063/1.4890441} {\bibfield  {journal} {\bibinfo  {journal} {The Journal of
  Chemical Physics}\ }\textbf {\bibinfo {volume} {141}},\ \bibinfo {pages}
  {044114} (\bibinfo {year} {2014})}\BibitemShut {NoStop}%
\bibitem [{\citenamefont {Tanimura}(2015)}]{T15JCP}%
  \BibitemOpen
  \bibfield  {author} {\bibinfo {author} {\bibfnamefont {Y.}~\bibnamefont
  {Tanimura}},\ }\bibfield  {title} {\enquote {\bibinfo {title} {Real-time and
  imaginary-time quantum hierarchal fokker-planck equations},}\ }\href
  {\doibase 10.1063/1.4916647} {\bibfield  {journal} {\bibinfo  {journal} {The
  Journal of Chemical Physics}\ }\textbf {\bibinfo {volume} {142}},\ \bibinfo
  {pages} {144110} (\bibinfo {year} {2015})}\BibitemShut {NoStop}%
\bibitem [{\citenamefont {Tanimura}(2020)}]{T20JCP}%
  \BibitemOpen
  \bibfield  {author} {\bibinfo {author} {\bibfnamefont {Y.}~\bibnamefont
  {Tanimura}},\ }\bibfield  {title} {\enquote {\bibinfo {title} {Numerically
  "exact" approach to open quantum dynamics: The hierarchical equations of
  motion (heom)},}\ }\href {\doibase 10.1063/5.0011599} {\bibfield  {journal}
  {\bibinfo  {journal} {The Journal of Chemical Physics}\ }\textbf {\bibinfo
  {volume} {153}},\ \bibinfo {pages} {020901} (\bibinfo {year}
  {2020})}\BibitemShut {NoStop}%
\bibitem [{\citenamefont {Sakamoto}\ and\ \citenamefont
  {Tanimura}(2021)}]{ST21JPSJ}%
  \BibitemOpen
  \bibfield  {author} {\bibinfo {author} {\bibfnamefont {S.}~\bibnamefont
  {Sakamoto}}\ and\ \bibinfo {author} {\bibfnamefont {Y.}~\bibnamefont
  {Tanimura}},\ }\bibfield  {title} {\enquote {\bibinfo {title} {Open quantum
  ddynamics theory for non-equilibrium work: Hierarchical equations of motion
  approach},}\ }\href {\doibase 10.7566/JPSJ.90.033001} {\bibfield  {journal}
  {\bibinfo  {journal} {Journal of the Physical Society of Japan}\ }\textbf
  {\bibinfo {volume} {90}},\ \bibinfo {pages} {033001} (\bibinfo {year}
  {2021})}\BibitemShut {NoStop}%
\bibitem [{\citenamefont {Sakamoto}\ and\ \citenamefont
  {Tanimura}(2020)}]{ST20JCP}%
  \BibitemOpen
  \bibfield  {author} {\bibinfo {author} {\bibfnamefont {S.}~\bibnamefont
  {Sakamoto}}\ and\ \bibinfo {author} {\bibfnamefont {Y.}~\bibnamefont
  {Tanimura}},\ }\bibfield  {title} {\enquote {\bibinfo {title} {Numerically
  "exact" simulations of entropy production in the fully quantum regime:
  Boltzmann entropy vs von neumann entropy},}\ }\href {\doibase
  10.1063/5.0033664} {\bibfield  {journal} {\bibinfo  {journal} {The Journal of
  Chemical Physics}\ }\textbf {\bibinfo {volume} {153}},\ \bibinfo {pages}
  {234107} (\bibinfo {year} {2020})}\BibitemShut {NoStop}%
\bibitem [{\citenamefont {Kato}\ and\ \citenamefont
  {Tanimura}(2015)}]{KT15JCP}%
  \BibitemOpen
  \bibfield  {author} {\bibinfo {author} {\bibfnamefont {A.}~\bibnamefont
  {Kato}}\ and\ \bibinfo {author} {\bibfnamefont {Y.}~\bibnamefont
  {Tanimura}},\ }\bibfield  {title} {\enquote {\bibinfo {title} {Quantum heat
  transport of a two-qubit system: Interplay between system-bath coherence and
  qubit-qubit coherence},}\ }\href {\doibase 10.1063/1.4928192} {\bibfield
  {journal} {\bibinfo  {journal} {The Journal of Chemical Physics}\ }\textbf
  {\bibinfo {volume} {143}},\ \bibinfo {pages} {064107} (\bibinfo {year}
  {2015})}\BibitemShut {NoStop}%
\bibitem [{\citenamefont {Kato}\ and\ \citenamefont
  {Tanimura}(2016)}]{KT16JCP}%
  \BibitemOpen
  \bibfield  {author} {\bibinfo {author} {\bibfnamefont {A.}~\bibnamefont
  {Kato}}\ and\ \bibinfo {author} {\bibfnamefont {Y.}~\bibnamefont
  {Tanimura}},\ }\bibfield  {title} {\enquote {\bibinfo {title} {Quantum heat
  current under non-perturbative and non-markovian conditions: Applications to
  heat machines},}\ }\href {\doibase 10.1063/1.4971370} {\bibfield  {journal}
  {\bibinfo  {journal} {The Journal of Chemical Physics}\ }\textbf {\bibinfo
  {volume} {145}},\ \bibinfo {pages} {224105} (\bibinfo {year}
  {2016})}\BibitemShut {NoStop}%
\bibitem [{\citenamefont {Song}\ and\ \citenamefont {Shi}(2017)}]{Shi2017}%
  \BibitemOpen
  \bibfield  {author} {\bibinfo {author} {\bibfnamefont {L.}~\bibnamefont
  {Song}}\ and\ \bibinfo {author} {\bibfnamefont {Q.}~\bibnamefont {Shi}},\
  }\bibfield  {title} {\enquote {\bibinfo {title} {Hierarchical equations of
  motion method applied to nonequilibrium heat transport in model molecular
  junctions: Transient heat current and high-order moments of the current
  operator},}\ }\href {\doibase 10.1103/PhysRevB.95.064308} {\bibfield
  {journal} {\bibinfo  {journal} {Phys. Rev. B}\ }\textbf {\bibinfo {volume}
  {95}},\ \bibinfo {pages} {064308} (\bibinfo {year} {2017})}\BibitemShut
  {NoStop}%
\bibitem [{\citenamefont {H\"artle}\ \emph {et~al.}(2013)\citenamefont
  {H\"artle}, \citenamefont {Cohen}, \citenamefont {Reichman},\ and\
  \citenamefont {Millis}}]{ReichmanPhysRevB2013}%
  \BibitemOpen
  \bibfield  {author} {\bibinfo {author} {\bibfnamefont {R.}~\bibnamefont
  {H\"artle}}, \bibinfo {author} {\bibfnamefont {G.}~\bibnamefont {Cohen}},
  \bibinfo {author} {\bibfnamefont {D.~R.}\ \bibnamefont {Reichman}}, \ and\
  \bibinfo {author} {\bibfnamefont {A.~J.}\ \bibnamefont {Millis}},\ }\bibfield
   {title} {\enquote {\bibinfo {title} {Decoherence and lead-induced interdot
  coupling in nonequilibrium electron transport through interacting quantum
  dots: A hierarchical quantum master equation approach},}\ }\href {\doibase
  10.1103/PhysRevB.88.235426} {\bibfield  {journal} {\bibinfo  {journal} {Phys.
  Rev. B}\ }\textbf {\bibinfo {volume} {88}},\ \bibinfo {pages} {235426}
  (\bibinfo {year} {2013})}\BibitemShut {NoStop}%
\bibitem [{\citenamefont {Goyal}, \citenamefont {He},\ and\ \citenamefont
  {Kawai}(2020)}]{GoyalKawai2020}%
  \BibitemOpen
  \bibfield  {author} {\bibinfo {author} {\bibfnamefont {K.}~\bibnamefont
  {Goyal}}, \bibinfo {author} {\bibfnamefont {X.}~\bibnamefont {He}}, \ and\
  \bibinfo {author} {\bibfnamefont {R.}~\bibnamefont {Kawai}},\ }\bibfield
  {title} {\enquote {\bibinfo {title} {Entropy production of a small quantum
  system under strong coupling with an environment: A computational
  experiment},}\ }\href {\doibase https://doi.org/10.1016/j.physa.2019.122627}
  {\bibfield  {journal} {\bibinfo  {journal} {Physica A: Statistical Mechanics
  and its Applications}\ }\textbf {\bibinfo {volume} {552}},\ \bibinfo {pages}
  {122627} (\bibinfo {year} {2020})},\ \bibinfo {note} {tributes of
  Non-equilibrium Statistical Physics}\BibitemShut {NoStop}%
\bibitem [{\citenamefont {Xu}, \citenamefont {Stockburger},\ and\ \citenamefont
  {Ankerhold}(2021)}]{Xu_Ankerhold2021}%
  \BibitemOpen
  \bibfield  {author} {\bibinfo {author} {\bibfnamefont {M.}~\bibnamefont
  {Xu}}, \bibinfo {author} {\bibfnamefont {J.~T.}\ \bibnamefont {Stockburger}},
  \ and\ \bibinfo {author} {\bibfnamefont {J.}~\bibnamefont {Ankerhold}},\
  }\bibfield  {title} {\enquote {\bibinfo {title} {Heat transport through a
  superconducting artificial atom},}\ }\href {\doibase
  10.1103/PhysRevB.103.104304} {\bibfield  {journal} {\bibinfo  {journal}
  {Phys. Rev. B}\ }\textbf {\bibinfo {volume} {103}},\ \bibinfo {pages}
  {104304} (\bibinfo {year} {2021})}\BibitemShut {NoStop}%
\bibitem [{\citenamefont {Quan}\ \emph {et~al.}(2007)\citenamefont {Quan},
  \citenamefont {Liu}, \citenamefont {Sun},\ and\ \citenamefont
  {Nori}}]{Nori2007}%
  \BibitemOpen
  \bibfield  {author} {\bibinfo {author} {\bibfnamefont {H.~T.}\ \bibnamefont
  {Quan}}, \bibinfo {author} {\bibfnamefont {Y.-x.}\ \bibnamefont {Liu}},
  \bibinfo {author} {\bibfnamefont {C.~P.}\ \bibnamefont {Sun}}, \ and\
  \bibinfo {author} {\bibfnamefont {F.}~\bibnamefont {Nori}},\ }\bibfield
  {title} {\enquote {\bibinfo {title} {Quantum thermodynamic cycles and quantum
  heat engines},}\ }\href {\doibase 10.1103/PhysRevE.76.031105} {\bibfield
  {journal} {\bibinfo  {journal} {Phys. Rev. E}\ }\textbf {\bibinfo {volume}
  {76}},\ \bibinfo {pages} {031105} (\bibinfo {year} {2007})}\BibitemShut
  {NoStop}%
\bibitem [{\citenamefont {Ono}\ \emph {et~al.}(2020)\citenamefont {Ono},
  \citenamefont {Shevchenko}, \citenamefont {Mori}, \citenamefont {Moriyama},\
  and\ \citenamefont {Nori}}]{OttoNori2020}%
  \BibitemOpen
  \bibfield  {author} {\bibinfo {author} {\bibfnamefont {K.}~\bibnamefont
  {Ono}}, \bibinfo {author} {\bibfnamefont {S.~N.}\ \bibnamefont {Shevchenko}},
  \bibinfo {author} {\bibfnamefont {T.}~\bibnamefont {Mori}}, \bibinfo {author}
  {\bibfnamefont {S.}~\bibnamefont {Moriyama}}, \ and\ \bibinfo {author}
  {\bibfnamefont {F.}~\bibnamefont {Nori}},\ }\bibfield  {title} {\enquote
  {\bibinfo {title} {Analog of a quantum heat engine using a single-spin
  qubit},}\ }\href {\doibase 10.1103/PhysRevLett.125.166802} {\bibfield
  {journal} {\bibinfo  {journal} {Phys. Rev. Lett.}\ }\textbf {\bibinfo
  {volume} {125}},\ \bibinfo {pages} {166802} (\bibinfo {year}
  {2020})}\BibitemShut {NoStop}%
\bibitem [{\citenamefont {Gardas}\ and\ \citenamefont
  {Deffner}(2015)}]{CarnotDeffner2015}%
  \BibitemOpen
  \bibfield  {author} {\bibinfo {author} {\bibfnamefont {B.}~\bibnamefont
  {Gardas}}\ and\ \bibinfo {author} {\bibfnamefont {S.}~\bibnamefont
  {Deffner}},\ }\bibfield  {title} {\enquote {\bibinfo {title} {Thermodynamic
  universality of quantum carnot engines},}\ }\href {\doibase
  10.1103/PhysRevE.92.042126} {\bibfield  {journal} {\bibinfo  {journal} {Phys.
  Rev. E}\ }\textbf {\bibinfo {volume} {92}},\ \bibinfo {pages} {042126}
  (\bibinfo {year} {2015})}\BibitemShut {NoStop}%
\bibitem [{\citenamefont {Kosloff}\ and\ \citenamefont
  {Levy}(2014)}]{Kosloff2014}%
  \BibitemOpen
  \bibfield  {author} {\bibinfo {author} {\bibfnamefont {R.}~\bibnamefont
  {Kosloff}}\ and\ \bibinfo {author} {\bibfnamefont {A.}~\bibnamefont {Levy}},\
  }\bibfield  {title} {\enquote {\bibinfo {title} {Quantum heat engines and
  refrigerators: Continuous devices},}\ }\href {\doibase
  10.1146/annurev-physchem-040513-103724} {\bibfield  {journal} {\bibinfo
  {journal} {Annual Review of Physical Chemistry}\ }\textbf {\bibinfo {volume}
  {65}},\ \bibinfo {pages} {365--393} (\bibinfo {year} {2014})},\ \bibinfo
  {note} {pMID: 24689798}\BibitemShut {NoStop}%
\bibitem [{\citenamefont {Kosloff}\ and\ \citenamefont
  {Rezek}(2017)}]{Kosloff2017}%
  \BibitemOpen
  \bibfield  {author} {\bibinfo {author} {\bibfnamefont {R.}~\bibnamefont
  {Kosloff}}\ and\ \bibinfo {author} {\bibfnamefont {Y.}~\bibnamefont
  {Rezek}},\ }\bibfield  {title} {\enquote {\bibinfo {title} {The quantum
  harmonic otto cycle},}\ }\href {\doibase 10.3390/e19040136} {\bibfield
  {journal} {\bibinfo  {journal} {Entropy}\ }\textbf {\bibinfo {volume} {19}}
  (\bibinfo {year} {2017}),\ 10.3390/e19040136}\BibitemShut {NoStop}%
\bibitem [{\citenamefont {Gelbwaser-Klimovsky}, \citenamefont {Niedenzu},\ and\
  \citenamefont {Kurizki}(2015)}]{KurizkiReview2015}%
  \BibitemOpen
  \bibfield  {author} {\bibinfo {author} {\bibfnamefont {D.}~\bibnamefont
  {Gelbwaser-Klimovsky}}, \bibinfo {author} {\bibfnamefont {W.}~\bibnamefont
  {Niedenzu}}, \ and\ \bibinfo {author} {\bibfnamefont {G.}~\bibnamefont
  {Kurizki}},\ }\bibfield  {title} {\enquote {\bibinfo {title} {Chapter twelve
  - thermodynamics of quantum systems under dynamical control},}\ \ }(\bibinfo
  {publisher} {Academic Press},\ \bibinfo {year} {2015})\ pp.\ \bibinfo {pages}
  {329--407}\BibitemShut {NoStop}%
\bibitem [{\citenamefont {Esposito}, \citenamefont {Ochoa},\ and\ \citenamefont
  {Galperin}(2015)}]{GalperinPhysRevB2015}%
  \BibitemOpen
  \bibfield  {author} {\bibinfo {author} {\bibfnamefont {M.}~\bibnamefont
  {Esposito}}, \bibinfo {author} {\bibfnamefont {M.~A.}\ \bibnamefont {Ochoa}},
  \ and\ \bibinfo {author} {\bibfnamefont {M.}~\bibnamefont {Galperin}},\
  }\bibfield  {title} {\enquote {\bibinfo {title} {Nature of heat in strongly
  coupled open quantum systems},}\ }\href {\doibase 10.1103/PhysRevB.92.235440}
  {\bibfield  {journal} {\bibinfo  {journal} {Phys. Rev. B}\ }\textbf {\bibinfo
  {volume} {92}},\ \bibinfo {pages} {235440} (\bibinfo {year}
  {2015})}\BibitemShut {NoStop}%
\bibitem [{\citenamefont {Seshadri}\ and\ \citenamefont
  {Galperin}(2021)}]{GalperinPhysRevB2021}%
  \BibitemOpen
  \bibfield  {author} {\bibinfo {author} {\bibfnamefont {N.}~\bibnamefont
  {Seshadri}}\ and\ \bibinfo {author} {\bibfnamefont {M.}~\bibnamefont
  {Galperin}},\ }\bibfield  {title} {\enquote {\bibinfo {title} {Entropy and
  information flow in quantum systems strongly coupled to baths},}\ }\href
  {\doibase 10.1103/PhysRevB.103.085415} {\bibfield  {journal} {\bibinfo
  {journal} {Phys. Rev. B}\ }\textbf {\bibinfo {volume} {103}},\ \bibinfo
  {pages} {085415} (\bibinfo {year} {2021})}\BibitemShut {NoStop}%
\bibitem [{\citenamefont {Aurell}(2017)}]{Aurell2017}%
  \BibitemOpen
  \bibfield  {author} {\bibinfo {author} {\bibfnamefont {E.}~\bibnamefont
  {Aurell}},\ }\bibfield  {title} {\enquote {\bibinfo {title} {On work and heat
  in time-dependent strong coupling},}\ }\href
  {https://www.mdpi.com/1099-4300/19/11/595} {\bibfield  {journal} {\bibinfo
  {journal} {Entropy}\ }\textbf {\bibinfo {volume} {19}} (\bibinfo {year}
  {2017})}\BibitemShut {NoStop}%
\bibitem [{\citenamefont {Xu}, \citenamefont {Chen},\ and\ \citenamefont
  {Liu}(2018)}]{Carnot_efficiency2018}%
  \BibitemOpen
  \bibfield  {author} {\bibinfo {author} {\bibfnamefont {Y.~Y.}\ \bibnamefont
  {Xu}}, \bibinfo {author} {\bibfnamefont {B.}~\bibnamefont {Chen}}, \ and\
  \bibinfo {author} {\bibfnamefont {J.}~\bibnamefont {Liu}},\ }\bibfield
  {title} {\enquote {\bibinfo {title} {Achieving the classical carnot
  efficiency in a strongly coupled quantum heat engine},}\ }\href {\doibase
  10.1103/PhysRevE.97.022130} {\bibfield  {journal} {\bibinfo  {journal} {Phys.
  Rev. E}\ }\textbf {\bibinfo {volume} {97}},\ \bibinfo {pages} {022130}
  (\bibinfo {year} {2018})}\BibitemShut {NoStop}%
\bibitem [{\citenamefont {Kilgour}\ and\ \citenamefont
  {Segal}(2018)}]{refrigeSegal2018}%
  \BibitemOpen
  \bibfield  {author} {\bibinfo {author} {\bibfnamefont {M.}~\bibnamefont
  {Kilgour}}\ and\ \bibinfo {author} {\bibfnamefont {D.}~\bibnamefont
  {Segal}},\ }\bibfield  {title} {\enquote {\bibinfo {title} {Coherence and
  decoherence in quantum absorption refrigerators},}\ }\href {\doibase
  10.1103/PhysRevE.98.012117} {\bibfield  {journal} {\bibinfo  {journal} {Phys.
  Rev. E}\ }\textbf {\bibinfo {volume} {98}},\ \bibinfo {pages} {012117}
  (\bibinfo {year} {2018})}\BibitemShut {NoStop}%
\bibitem [{\citenamefont {Lee}\ \emph {et~al.}(2020)\citenamefont {Lee},
  \citenamefont {Ha}, \citenamefont {Park},\ and\ \citenamefont
  {Jeong}}]{PhysRevEOtto2020}%
  \BibitemOpen
  \bibfield  {author} {\bibinfo {author} {\bibfnamefont {S.}~\bibnamefont
  {Lee}}, \bibinfo {author} {\bibfnamefont {M.}~\bibnamefont {Ha}}, \bibinfo
  {author} {\bibfnamefont {J.-M.}\ \bibnamefont {Park}}, \ and\ \bibinfo
  {author} {\bibfnamefont {H.}~\bibnamefont {Jeong}},\ }\bibfield  {title}
  {\enquote {\bibinfo {title} {Finite-time quantum otto engine: Surpassing the
  quasistatic efficiency due to friction},}\ }\href {\doibase
  10.1103/PhysRevE.101.022127} {\bibfield  {journal} {\bibinfo  {journal}
  {Phys. Rev. E}\ }\textbf {\bibinfo {volume} {101}},\ \bibinfo {pages}
  {022127} (\bibinfo {year} {2020})}\BibitemShut {NoStop}%
\bibitem [{\citenamefont {Das}\ and\ \citenamefont
  {Mukherjee}(2020)}]{2020OttoVictor}%
  \BibitemOpen
  \bibfield  {author} {\bibinfo {author} {\bibfnamefont {A.}~\bibnamefont
  {Das}}\ and\ \bibinfo {author} {\bibfnamefont {V.}~\bibnamefont
  {Mukherjee}},\ }\bibfield  {title} {\enquote {\bibinfo {title}
  {Quantum-enhanced finite-time otto cycle},}\ }\href {\doibase
  10.1103/physrevresearch.2.033083} {\bibfield  {journal} {\bibinfo  {journal}
  {Physical Review Research}\ }\textbf {\bibinfo {volume} {2}} (\bibinfo {year}
  {2020}),\ 10.1103/physrevresearch.2.033083}\BibitemShut {NoStop}%
\bibitem [{\citenamefont {Johal}\ and\ \citenamefont
  {Mehta}(2021)}]{Otto2021Venu}%
  \BibitemOpen
  \bibfield  {author} {\bibinfo {author} {\bibfnamefont {R.~S.}\ \bibnamefont
  {Johal}}\ and\ \bibinfo {author} {\bibfnamefont {V.}~\bibnamefont {Mehta}},\
  }\bibfield  {title} {\enquote {\bibinfo {title} {Quantum heat engines with
  complex working media, complete otto cycles and heuristics},}\ }\href
  {\doibase 10.3390/e23091149} {\bibfield  {journal} {\bibinfo  {journal}
  {Entropy}\ }\textbf {\bibinfo {volume} {23}} (\bibinfo {year} {2021}),\
  10.3390/e23091149}\BibitemShut {NoStop}%
\bibitem [{\citenamefont {Pezzutto}, \citenamefont {Paternostro},\ and\
  \citenamefont {Omar}(2019)}]{non-MarkovianPezzutto_2019}%
  \BibitemOpen
  \bibfield  {author} {\bibinfo {author} {\bibfnamefont {M.}~\bibnamefont
  {Pezzutto}}, \bibinfo {author} {\bibfnamefont {M.}~\bibnamefont
  {Paternostro}}, \ and\ \bibinfo {author} {\bibfnamefont {Y.}~\bibnamefont
  {Omar}},\ }\bibfield  {title} {\enquote {\bibinfo {title} {An
  out-of-equilibrium non-markovian quantum heat engine},}\ }\href {\doibase
  10.1088/2058-9565/aaf5b4} {\bibfield  {journal} {\bibinfo  {journal} {Quantum
  Science and Technology}\ }\textbf {\bibinfo {volume} {4}},\ \bibinfo {pages}
  {025002} (\bibinfo {year} {2019})}\BibitemShut {NoStop}%
\bibitem [{\citenamefont {Abiuso}\ and\ \citenamefont
  {Giovannetti}(2019)}]{Non-MarkovGiovannetti2019}%
  \BibitemOpen
  \bibfield  {author} {\bibinfo {author} {\bibfnamefont {P.}~\bibnamefont
  {Abiuso}}\ and\ \bibinfo {author} {\bibfnamefont {V.}~\bibnamefont
  {Giovannetti}},\ }\bibfield  {title} {\enquote {\bibinfo {title} {Non-markov
  enhancement of maximum power for quantum thermal machines},}\ }\href
  {\doibase 10.1103/PhysRevA.99.052106} {\bibfield  {journal} {\bibinfo
  {journal} {Phys. Rev. A}\ }\textbf {\bibinfo {volume} {99}},\ \bibinfo
  {pages} {052106} (\bibinfo {year} {2019})}\BibitemShut {NoStop}%
\bibitem [{\citenamefont {Wiedmann}, \citenamefont {Stockburger},\ and\
  \citenamefont {Ankerhold}(2020)}]{Non-MarkovWiedmann_Ankerhold2020}%
  \BibitemOpen
  \bibfield  {author} {\bibinfo {author} {\bibfnamefont {M.}~\bibnamefont
  {Wiedmann}}, \bibinfo {author} {\bibfnamefont {J.~T.}\ \bibnamefont
  {Stockburger}}, \ and\ \bibinfo {author} {\bibfnamefont {J.}~\bibnamefont
  {Ankerhold}},\ }\bibfield  {title} {\enquote {\bibinfo {title} {Non-markovian
  dynamics of a quantum heat engine: out-of-equilibrium operation and thermal
  coupling control},}\ }\href {\doibase 10.1088/1367-2630/ab725a} {\bibfield
  {journal} {\bibinfo  {journal} {New Journal of Physics}\ }\textbf {\bibinfo
  {volume} {22}},\ \bibinfo {pages} {033007} (\bibinfo {year}
  {2020})}\BibitemShut {NoStop}%
\bibitem [{\citenamefont {Wiedmann}, \citenamefont {Stockburger},\ and\
  \citenamefont {Ankerhold}(2021)}]{OttoWiedmann_Ankerhold2021}%
  \BibitemOpen
  \bibfield  {author} {\bibinfo {author} {\bibfnamefont {M.}~\bibnamefont
  {Wiedmann}}, \bibinfo {author} {\bibfnamefont {J.~T.}\ \bibnamefont
  {Stockburger}}, \ and\ \bibinfo {author} {\bibfnamefont {J.}~\bibnamefont
  {Ankerhold}},\ }\bibfield  {title} {\enquote {\bibinfo {title} {Non-markovian
  quantum otto refrigerator},}\ }\href {\doibase
  10.1140/epjs/s11734-021-00094-0} {\bibfield  {journal} {\bibinfo  {journal}
  {Eur. Phys. J. Spec. Tops}\ }\textbf {\bibinfo {volume} {230}},\ \bibinfo
  {pages} {851--857} (\bibinfo {year} {2021})}\BibitemShut {NoStop}%
\bibitem [{\citenamefont {Xu}\ \emph {et~al.}(2022)\citenamefont {Xu},
  \citenamefont {Stockburger}, \citenamefont {Kurizki},\ and\ \citenamefont
  {Ankerhold}}]{Xu_Ankerhold2022}%
  \BibitemOpen
  \bibfield  {author} {\bibinfo {author} {\bibfnamefont {M.}~\bibnamefont
  {Xu}}, \bibinfo {author} {\bibfnamefont {J.~T.}\ \bibnamefont {Stockburger}},
  \bibinfo {author} {\bibfnamefont {G.}~\bibnamefont {Kurizki}}, \ and\
  \bibinfo {author} {\bibfnamefont {J.}~\bibnamefont {Ankerhold}},\ }\bibfield
  {title} {\enquote {\bibinfo {title} {Minimal quantum thermal machine in a
  bandgap environment: non-markovian features and anti-zeno advantage},}\
  }\href {\doibase 10.1088/1367-2630/ac575b} {\bibfield  {journal} {\bibinfo
  {journal} {New Journal of Physics}\ }\textbf {\bibinfo {volume} {24}},\
  \bibinfo {pages} {035003} (\bibinfo {year} {2022})}\BibitemShut {NoStop}%
\bibitem [{\citenamefont {Brenes}\ \emph {et~al.}(2020)\citenamefont {Brenes},
  \citenamefont {Mendoza-Arenas}, \citenamefont {Purkayastha}, \citenamefont
  {Mitchison}, \citenamefont {Clark},\ and\ \citenamefont
  {Goold}}]{TensorNetPhysRevX2020}%
  \BibitemOpen
  \bibfield  {author} {\bibinfo {author} {\bibfnamefont {M.}~\bibnamefont
  {Brenes}}, \bibinfo {author} {\bibfnamefont {J.~J.}\ \bibnamefont
  {Mendoza-Arenas}}, \bibinfo {author} {\bibfnamefont {A.}~\bibnamefont
  {Purkayastha}}, \bibinfo {author} {\bibfnamefont {M.~T.}\ \bibnamefont
  {Mitchison}}, \bibinfo {author} {\bibfnamefont {S.~R.}\ \bibnamefont
  {Clark}}, \ and\ \bibinfo {author} {\bibfnamefont {J.}~\bibnamefont
  {Goold}},\ }\bibfield  {title} {\enquote {\bibinfo {title} {Tensor-network
  method to simulate strongly interacting quantum thermal machines},}\ }\href
  {\doibase 10.1103/PhysRevX.10.031040} {\bibfield  {journal} {\bibinfo
  {journal} {Phys. Rev. X}\ }\textbf {\bibinfo {volume} {10}},\ \bibinfo
  {pages} {031040} (\bibinfo {year} {2020})}\BibitemShut {NoStop}%
\bibitem [{\citenamefont {Pancotti}\ \emph {et~al.}(2020)\citenamefont
  {Pancotti}, \citenamefont {Scandi}, \citenamefont {Mitchison},\ and\
  \citenamefont {Perarnau-Llobet}}]{PhysRev2020SBcouple}%
  \BibitemOpen
  \bibfield  {author} {\bibinfo {author} {\bibfnamefont {N.}~\bibnamefont
  {Pancotti}}, \bibinfo {author} {\bibfnamefont {M.}~\bibnamefont {Scandi}},
  \bibinfo {author} {\bibfnamefont {M.~T.}\ \bibnamefont {Mitchison}}, \ and\
  \bibinfo {author} {\bibfnamefont {M.}~\bibnamefont {Perarnau-Llobet}},\
  }\bibfield  {title} {\enquote {\bibinfo {title} {Speed-ups to isothermality:
  Enhanced quantum thermal machines through control of the system-bath
  coupling},}\ }\href {\doibase 10.1103/PhysRevX.10.031015} {\bibfield
  {journal} {\bibinfo  {journal} {Phys. Rev. X}\ }\textbf {\bibinfo {volume}
  {10}},\ \bibinfo {pages} {031015} (\bibinfo {year} {2020})}\BibitemShut
  {NoStop}%
\bibitem [{\citenamefont {Cangemi}\ \emph {et~al.}(2020)\citenamefont
  {Cangemi}, \citenamefont {Cataudella}, \citenamefont {Benenti}, \citenamefont
  {Sassetti},\ and\ \citenamefont {De~Filippis}}]{PhysRevBUncertenty2020}%
  \BibitemOpen
  \bibfield  {author} {\bibinfo {author} {\bibfnamefont {L.~M.}\ \bibnamefont
  {Cangemi}}, \bibinfo {author} {\bibfnamefont {V.}~\bibnamefont {Cataudella}},
  \bibinfo {author} {\bibfnamefont {G.}~\bibnamefont {Benenti}}, \bibinfo
  {author} {\bibfnamefont {M.}~\bibnamefont {Sassetti}}, \ and\ \bibinfo
  {author} {\bibfnamefont {G.}~\bibnamefont {De~Filippis}},\ }\bibfield
  {title} {\enquote {\bibinfo {title} {Violation of thermodynamics uncertainty
  relations in a periodically driven work-to-work converter from weak to strong
  dissipation},}\ }\href {\doibase 10.1103/PhysRevB.102.165418} {\bibfield
  {journal} {\bibinfo  {journal} {Phys. Rev. B}\ }\textbf {\bibinfo {volume}
  {102}},\ \bibinfo {pages} {165418} (\bibinfo {year} {2020})}\BibitemShut
  {NoStop}%
\bibitem [{\citenamefont {Clapeyron}(1834)}]{Clapeyron1834}%
  \BibitemOpen
  \bibfield  {author} {\bibinfo {author} {\bibfnamefont {B.~P.~E.}\
  \bibnamefont {Clapeyron}},\ }\bibfield  {title} {\enquote {\bibinfo {title}
  {Memoire sur la puissance motrice de la chaleur},}\ }\href@noop {} {\bibfield
   {journal} {\bibinfo  {journal} {Journal de l' Ecole Royale Polytechnique}\
  }\textbf {\bibinfo {volume} {14}},\ \bibinfo {pages} {153--190} (\bibinfo
  {year} {1834})}\BibitemShut {NoStop}%
\bibitem [{\citenamefont {An}\ \emph {et~al.}(2015)\citenamefont {An},
  \citenamefont {Zhang}, \citenamefont {Um}, \citenamefont {Lv}, \citenamefont
  {Lu}, \citenamefont {Zhang}, \citenamefont {Yin}, \citenamefont {Quan},\ and\
  \citenamefont {Kim}}]{JarEXE2015}%
  \BibitemOpen
  \bibfield  {author} {\bibinfo {author} {\bibfnamefont {S.}~\bibnamefont
  {An}}, \bibinfo {author} {\bibfnamefont {J.-N.}\ \bibnamefont {Zhang}},
  \bibinfo {author} {\bibfnamefont {M.}~\bibnamefont {Um}}, \bibinfo {author}
  {\bibfnamefont {D.}~\bibnamefont {Lv}}, \bibinfo {author} {\bibfnamefont
  {Y.}~\bibnamefont {Lu}}, \bibinfo {author} {\bibfnamefont {J.}~\bibnamefont
  {Zhang}}, \bibinfo {author} {\bibfnamefont {Z.-Q.}\ \bibnamefont {Yin}},
  \bibinfo {author} {\bibfnamefont {H.~T.}\ \bibnamefont {Quan}}, \ and\
  \bibinfo {author} {\bibfnamefont {K.}~\bibnamefont {Kim}},\ }\bibfield
  {title} {\enquote {\bibinfo {title} {Experimental test of the quantum
  jarzynski equality with a trapped-ion system},}\ }\href {\doibase
  10.1038/nphys3197} {\bibfield  {journal} {\bibinfo  {journal} {Nature
  Physics}\ }\textbf {\bibinfo {volume} {11}},\ \bibinfo {pages} {193}
  (\bibinfo {year} {2015})}\BibitemShut {NoStop}%
\bibitem [{\citenamefont {An}\ \emph {et~al.}(2016)\citenamefont {An},
  \citenamefont {Lv}, \citenamefont {del Campo~Adolfo},\ and\ \citenamefont
  {Kim}}]{AdiaEXE2016}%
  \BibitemOpen
  \bibfield  {author} {\bibinfo {author} {\bibfnamefont {S.}~\bibnamefont
  {An}}, \bibinfo {author} {\bibfnamefont {D.}~\bibnamefont {Lv}}, \bibinfo
  {author} {\bibnamefont {del Campo~Adolfo}}, \ and\ \bibinfo {author}
  {\bibfnamefont {K.}~\bibnamefont {Kim}},\ }\bibfield  {title} {\enquote
  {\bibinfo {title} {Shortcuts to adiabaticity by counterdiabatic driving for
  trapped-ion displacement in phase space},}\ }\href {\doibase
  10.1038/ncomms12999} {\bibfield  {journal} {\bibinfo  {journal} {Nature
  Communications}\ }\textbf {\bibinfo {volume} {7}},\ \bibinfo {pages} {12999}
  (\bibinfo {year} {2016})}\BibitemShut {NoStop}%
\bibitem [{\citenamefont {Vitanov}\ \emph {et~al.}(2017)\citenamefont
  {Vitanov}, \citenamefont {Rangelov}, \citenamefont {Shore},\ and\
  \citenamefont {Bergmann}}]{KlassRMP2017}%
  \BibitemOpen
  \bibfield  {author} {\bibinfo {author} {\bibfnamefont {N.~V.}\ \bibnamefont
  {Vitanov}}, \bibinfo {author} {\bibfnamefont {A.~A.}\ \bibnamefont
  {Rangelov}}, \bibinfo {author} {\bibfnamefont {B.~W.}\ \bibnamefont {Shore}},
  \ and\ \bibinfo {author} {\bibfnamefont {K.}~\bibnamefont {Bergmann}},\
  }\bibfield  {title} {\enquote {\bibinfo {title} {Stimulated raman adiabatic
  passage in physics, chemistry, and beyond},}\ }\href {\doibase
  10.1103/RevModPhys.89.015006} {\bibfield  {journal} {\bibinfo  {journal}
  {Rev. Mod. Phys.}\ }\textbf {\bibinfo {volume} {89}},\ \bibinfo {pages}
  {015006} (\bibinfo {year} {2017})}\BibitemShut {NoStop}%
\bibitem [{\citenamefont {Hoffmann}\ \emph {et~al.}(2018)\citenamefont
  {Hoffmann}, \citenamefont {Fahlvik}, \citenamefont {Thelander}, \citenamefont
  {Leijnse},\ and\ \citenamefont {Linke}}]{LundJosefsson2018}%
  \BibitemOpen
  \bibfield  {author} {\bibinfo {author} {\bibfnamefont {E.~A.}\ \bibnamefont
  {Hoffmann}}, \bibinfo {author} {\bibfnamefont {S.}~\bibnamefont {Fahlvik}},
  \bibinfo {author} {\bibfnamefont {C.}~\bibnamefont {Thelander}}, \bibinfo
  {author} {\bibfnamefont {M.}~\bibnamefont {Leijnse}}, \ and\ \bibinfo
  {author} {\bibfnamefont {H.}~\bibnamefont {Linke}},\ }\bibfield  {title}
  {\enquote {\bibinfo {title} {A quantum-dot heat engine operating close to the
  thermodynamic efficiency limits},}\ }\href {\doibase
  10.1038/s41565-018-0200-5} {\bibfield  {journal} {\bibinfo  {journal} {Nature
  Nanotechnology}\ }\textbf {\bibinfo {volume} {13}},\ \bibinfo {pages}
  {920--924} (\bibinfo {year} {2018})}\BibitemShut {NoStop}%
\bibitem [{\citenamefont {Bengtsson}\ \emph {et~al.}(2018)\citenamefont
  {Bengtsson}, \citenamefont {Tengstrand}, \citenamefont {Wacker},
  \citenamefont {Samuelsson}, \citenamefont {Ueda}, \citenamefont {Linke},\
  and\ \citenamefont {Reimann}}]{ReimannPRL2018}%
  \BibitemOpen
  \bibfield  {author} {\bibinfo {author} {\bibfnamefont {J.}~\bibnamefont
  {Bengtsson}}, \bibinfo {author} {\bibfnamefont {M.~N.}\ \bibnamefont
  {Tengstrand}}, \bibinfo {author} {\bibfnamefont {A.}~\bibnamefont {Wacker}},
  \bibinfo {author} {\bibfnamefont {P.}~\bibnamefont {Samuelsson}}, \bibinfo
  {author} {\bibfnamefont {M.}~\bibnamefont {Ueda}}, \bibinfo {author}
  {\bibfnamefont {H.}~\bibnamefont {Linke}}, \ and\ \bibinfo {author}
  {\bibfnamefont {S.~M.}\ \bibnamefont {Reimann}},\ }\bibfield  {title}
  {\enquote {\bibinfo {title} {Quantum szilard engine with attractively
  interacting bosons},}\ }\href {\doibase 10.1103/PhysRevLett.120.100601}
  {\bibfield  {journal} {\bibinfo  {journal} {Phys. Rev. Lett.}\ }\textbf
  {\bibinfo {volume} {120}},\ \bibinfo {pages} {100601} (\bibinfo {year}
  {2018})}\BibitemShut {NoStop}%
\bibitem [{\citenamefont {Josefsson}\ \emph {et~al.}(2019)\citenamefont
  {Josefsson}, \citenamefont {Svilans}, \citenamefont {Linke},\ and\
  \citenamefont {Leijnse}}]{LundJosefsson2019}%
  \BibitemOpen
  \bibfield  {author} {\bibinfo {author} {\bibfnamefont {M.}~\bibnamefont
  {Josefsson}}, \bibinfo {author} {\bibfnamefont {A.}~\bibnamefont {Svilans}},
  \bibinfo {author} {\bibfnamefont {H.}~\bibnamefont {Linke}}, \ and\ \bibinfo
  {author} {\bibfnamefont {M.}~\bibnamefont {Leijnse}},\ }\bibfield  {title}
  {\enquote {\bibinfo {title} {Optimal power and efficiency of single quantum
  dot heat engines: Theory and experiment},}\ }\href {\doibase
  10.1103/PhysRevB.99.235432} {\bibfield  {journal} {\bibinfo  {journal} {Phys.
  Rev. B}\ }\textbf {\bibinfo {volume} {99}},\ \bibinfo {pages} {235432}
  (\bibinfo {year} {2019})}\BibitemShut {NoStop}%
\bibitem [{\citenamefont {Prete}\ \emph {et~al.}(2019)\citenamefont {Prete},
  \citenamefont {Erdman}, \citenamefont {Demontis}, \citenamefont {Zannier},
  \citenamefont {Ercolani}, \citenamefont {Sorba}, \citenamefont {Beltram},
  \citenamefont {Rossella}, \citenamefont {Taddei},\ and\ \citenamefont
  {Roddaro}}]{nanolettDomenic2019}%
  \BibitemOpen
  \bibfield  {author} {\bibinfo {author} {\bibfnamefont {D.}~\bibnamefont
  {Prete}}, \bibinfo {author} {\bibfnamefont {P.~A.}\ \bibnamefont {Erdman}},
  \bibinfo {author} {\bibfnamefont {V.}~\bibnamefont {Demontis}}, \bibinfo
  {author} {\bibfnamefont {V.}~\bibnamefont {Zannier}}, \bibinfo {author}
  {\bibfnamefont {D.}~\bibnamefont {Ercolani}}, \bibinfo {author}
  {\bibfnamefont {L.}~\bibnamefont {Sorba}}, \bibinfo {author} {\bibfnamefont
  {F.}~\bibnamefont {Beltram}}, \bibinfo {author} {\bibfnamefont
  {F.}~\bibnamefont {Rossella}}, \bibinfo {author} {\bibfnamefont
  {F.}~\bibnamefont {Taddei}}, \ and\ \bibinfo {author} {\bibfnamefont
  {S.}~\bibnamefont {Roddaro}},\ }\bibfield  {title} {\enquote {\bibinfo
  {title} {Thermoelectric conversion at 30 k in inas/inp nanowire quantum
  dots},}\ }\href {\doibase 10.1021/acs.nanolett.9b00276} {\bibfield  {journal}
  {\bibinfo  {journal} {Nano Letters}\ }\textbf {\bibinfo {volume} {19}},\
  \bibinfo {pages} {3033--3039} (\bibinfo {year} {2019})},\ \bibinfo {note}
  {pMID: 30935206}\BibitemShut {NoStop}%
\bibitem [{\citenamefont {Ma}\ \emph {et~al.}(2020)\citenamefont {Ma},
  \citenamefont {Zhai}, \citenamefont {Chen}, \citenamefont {Sun},\ and\
  \citenamefont {Dong}}]{HuiPRL2020}%
  \BibitemOpen
  \bibfield  {author} {\bibinfo {author} {\bibfnamefont {Y.-H.}\ \bibnamefont
  {Ma}}, \bibinfo {author} {\bibfnamefont {R.-X.}\ \bibnamefont {Zhai}},
  \bibinfo {author} {\bibfnamefont {J.}~\bibnamefont {Chen}}, \bibinfo {author}
  {\bibfnamefont {C.~P.}\ \bibnamefont {Sun}}, \ and\ \bibinfo {author}
  {\bibfnamefont {H.}~\bibnamefont {Dong}},\ }\bibfield  {title} {\enquote
  {\bibinfo {title} {Experimental test of the $1/\ensuremath{\tau}$-scaling
  entropy generation in finite-time thermodynamics},}\ }\href {\doibase
  10.1103/PhysRevLett.125.210601} {\bibfield  {journal} {\bibinfo  {journal}
  {Phys. Rev. Lett.}\ }\textbf {\bibinfo {volume} {125}},\ \bibinfo {pages}
  {210601} (\bibinfo {year} {2020})}\BibitemShut {NoStop}%
\bibitem [{\citenamefont {{de Oliveira}}\ \emph {et~al.}(2020)\citenamefont
  {{de Oliveira}}, \citenamefont {Gomes}, \citenamefont {Brasil}, \citenamefont
  {{Rubiano da Silva}}, \citenamefont {Céleri},\ and\ \citenamefont {{Souto
  Ribeiro}}}]{DEOLIVEIRA2020}%
  \BibitemOpen
  \bibfield  {author} {\bibinfo {author} {\bibfnamefont {A.}~\bibnamefont {{de
  Oliveira}}}, \bibinfo {author} {\bibfnamefont {R.}~\bibnamefont {Gomes}},
  \bibinfo {author} {\bibfnamefont {V.}~\bibnamefont {Brasil}}, \bibinfo
  {author} {\bibfnamefont {N.}~\bibnamefont {{Rubiano da Silva}}}, \bibinfo
  {author} {\bibfnamefont {L.}~\bibnamefont {Céleri}}, \ and\ \bibinfo
  {author} {\bibfnamefont {P.}~\bibnamefont {{Souto Ribeiro}}},\ }\bibfield
  {title} {\enquote {\bibinfo {title} {Full thermalization of a photonic
  qubit},}\ }\href {\doibase https://doi.org/10.1016/j.physleta.2020.126933}
  {\bibfield  {journal} {\bibinfo  {journal} {Physics Letters A}\ }\textbf
  {\bibinfo {volume} {384}},\ \bibinfo {pages} {126933} (\bibinfo {year}
  {2020})}\BibitemShut {NoStop}%
\bibitem [{\citenamefont {Hern{\'{a}}ndez-G{\'{o}}mez}\ \emph
  {et~al.}(2021)\citenamefont {Hern{\'{a}}ndez-G{\'{o}}mez}, \citenamefont
  {Staudenmaier}, \citenamefont {Campisi},\ and\ \citenamefont
  {Fabbri}}]{Hern_ndez_G_mez_2021}%
  \BibitemOpen
  \bibfield  {author} {\bibinfo {author} {\bibfnamefont {S.}~\bibnamefont
  {Hern{\'{a}}ndez-G{\'{o}}mez}}, \bibinfo {author} {\bibfnamefont
  {N.}~\bibnamefont {Staudenmaier}}, \bibinfo {author} {\bibfnamefont
  {M.}~\bibnamefont {Campisi}}, \ and\ \bibinfo {author} {\bibfnamefont
  {N.}~\bibnamefont {Fabbri}},\ }\bibfield  {title} {\enquote {\bibinfo {title}
  {Experimental test of fluctuation relations for driven open quantum systems
  with an {NV} center},}\ }\href {\doibase 10.1088/1367-2630/abfc6a} {\bibfield
   {journal} {\bibinfo  {journal} {New Journal of Physics}\ }\textbf {\bibinfo
  {volume} {23}},\ \bibinfo {pages} {065004} (\bibinfo {year}
  {2021})}\BibitemShut {NoStop}%
\bibitem [{\citenamefont {Thomson}(1851)}]{Thomson1851}%
  \BibitemOpen
  \bibfield  {author} {\bibinfo {author} {\bibfnamefont {W.}~\bibnamefont
  {Thomson}},\ }\bibfield  {title} {\enquote {\bibinfo {title} {On the
  dynamical theory of heat, with numerical results deduced from mr joule's
  equivalent of a thermal unit, and m. regnault's observations on steam},}\
  }\href@noop {} {\bibfield  {journal} {\bibinfo  {journal} {Transactions of
  the Royal Society of Edinburgh XX (part II)}\ ,\ \bibinfo {pages}
  {261–298}} (\bibinfo {year} {1851})}\BibitemShut {NoStop}%
\bibitem [{\citenamefont {Lenard}(1978)}]{Lenard1978}%
  \BibitemOpen
  \bibfield  {author} {\bibinfo {author} {\bibfnamefont {A.}~\bibnamefont
  {Lenard}},\ }\bibfield  {title} {\enquote {\bibinfo {title} {Thermodynamical
  proof of the gibbs formula for elementary quantum systems},}\ }\href
  {\doibase 10.1007/BF01011769} {\bibfield  {journal} {\bibinfo  {journal}
  {Journal of Statistical Physics}\ }\textbf {\bibinfo {volume} {19}},\
  \bibinfo {pages} {575} (\bibinfo {year} {1978})}\BibitemShut {NoStop}%
\bibitem [{\citenamefont {Tanimura}\ and\ \citenamefont
  {Kubo}(1989)}]{TK89JPSJ1}%
  \BibitemOpen
  \bibfield  {author} {\bibinfo {author} {\bibfnamefont {Y.}~\bibnamefont
  {Tanimura}}\ and\ \bibinfo {author} {\bibfnamefont {R.}~\bibnamefont
  {Kubo}},\ }\bibfield  {title} {\enquote {\bibinfo {title} {Time evolution of
  a quantum system in contact with a nearly gaussian-markoffian noise bath},}\
  }\href {\doibase 10.1143/JPSJ.58.101} {\bibfield  {journal} {\bibinfo
  {journal} {Journal of the Physical Society of Japan}\ }\textbf {\bibinfo
  {volume} {58}},\ \bibinfo {pages} {101--114} (\bibinfo {year}
  {1989})}\BibitemShut {NoStop}%
\bibitem [{\citenamefont {Tanimura}(1990)}]{T90PRA}%
  \BibitemOpen
  \bibfield  {author} {\bibinfo {author} {\bibfnamefont {Y.}~\bibnamefont
  {Tanimura}},\ }\bibfield  {title} {\enquote {\bibinfo {title}
  {Nonperturbative expansion method for a quantum system coupled to a
  harmonic-oscillator bath},}\ }\href {\doibase 10.1103/PhysRevA.41.6676}
  {\bibfield  {journal} {\bibinfo  {journal} {Phys. Rev. A}\ }\textbf {\bibinfo
  {volume} {41}},\ \bibinfo {pages} {6676--6687} (\bibinfo {year}
  {1990})}\BibitemShut {NoStop}%
\bibitem [{\citenamefont {Ishizaki}\ and\ \citenamefont
  {Tanimura}(2005)}]{IT05JPSJ}%
  \BibitemOpen
  \bibfield  {author} {\bibinfo {author} {\bibfnamefont {A.}~\bibnamefont
  {Ishizaki}}\ and\ \bibinfo {author} {\bibfnamefont {Y.}~\bibnamefont
  {Tanimura}},\ }\bibfield  {title} {\enquote {\bibinfo {title} {Quantum
  dynamics of system strongly coupled to low-temperature colored noise bath:
  Reduced hierarchy equations approach},}\ }\href {\doibase
  10.1143/JPSJ.74.3131} {\bibfield  {journal} {\bibinfo  {journal} {Journal of
  the Physical Society of Japan}\ }\textbf {\bibinfo {volume} {74}},\ \bibinfo
  {pages} {3131--3134} (\bibinfo {year} {2005})}\BibitemShut {NoStop}%
\bibitem [{\citenamefont {Hu}, \citenamefont {Xu},\ and\ \citenamefont
  {Yan}(2010)}]{hu2010communication}%
  \BibitemOpen
  \bibfield  {author} {\bibinfo {author} {\bibfnamefont {J.}~\bibnamefont
  {Hu}}, \bibinfo {author} {\bibfnamefont {R.-X.}\ \bibnamefont {Xu}}, \ and\
  \bibinfo {author} {\bibfnamefont {Y.}~\bibnamefont {Yan}},\ }\href@noop {}
  {\enquote {\bibinfo {title} {Communication: Pad{\'e} spectrum decomposition
  of fermi function and bose function},}\ } (\bibinfo {year}
  {2010})\BibitemShut {NoStop}%
\bibitem [{\citenamefont {Quan}\ \emph {et~al.}(2006)\citenamefont {Quan},
  \citenamefont {Wang}, \citenamefont {Liu}, \citenamefont {Sun},\ and\
  \citenamefont {Nori}}]{Maxwelldemon2006}%
  \BibitemOpen
  \bibfield  {author} {\bibinfo {author} {\bibfnamefont {H.~T.}\ \bibnamefont
  {Quan}}, \bibinfo {author} {\bibfnamefont {Y.~D.}\ \bibnamefont {Wang}},
  \bibinfo {author} {\bibfnamefont {Y.-x.}\ \bibnamefont {Liu}}, \bibinfo
  {author} {\bibfnamefont {C.~P.}\ \bibnamefont {Sun}}, \ and\ \bibinfo
  {author} {\bibfnamefont {F.}~\bibnamefont {Nori}},\ }\bibfield  {title}
  {\enquote {\bibinfo {title} {Maxwell's demon assisted thermodynamic cycle in
  superconducting quantum circuits},}\ }\href {\doibase
  10.1103/PhysRevLett.97.180402} {\bibfield  {journal} {\bibinfo  {journal}
  {Phys. Rev. Lett.}\ }\textbf {\bibinfo {volume} {97}},\ \bibinfo {pages}
  {180402} (\bibinfo {year} {2006})}\BibitemShut {NoStop}%
\bibitem [{\citenamefont {Maruyama}, \citenamefont {Nori},\ and\ \citenamefont
  {Vedral}(2009)}]{Maxwelldemon2009}%
  \BibitemOpen
  \bibfield  {author} {\bibinfo {author} {\bibfnamefont {K.}~\bibnamefont
  {Maruyama}}, \bibinfo {author} {\bibfnamefont {F.}~\bibnamefont {Nori}}, \
  and\ \bibinfo {author} {\bibfnamefont {V.}~\bibnamefont {Vedral}},\
  }\bibfield  {title} {\enquote {\bibinfo {title} {Colloquium: The physics of
  maxwell's demon and information},}\ }\href {\doibase 10.1103/RevModPhys.81.1}
  {\bibfield  {journal} {\bibinfo  {journal} {Rev. Mod. Phys.}\ }\textbf
  {\bibinfo {volume} {81}},\ \bibinfo {pages} {1--23} (\bibinfo {year}
  {2009})}\BibitemShut {NoStop}%
\end{thebibliography}%
%%%%%%%%%%%%
\end{document}